\documentstyle[aas2pp4,psfig]{article}

\newcommand{\Vec}[1]{{\mbox{\boldmath{${#1}$}}}}
\begin{document}
\lefthead{Kawai {\it et al.}}
\righthead{GRAPE-5}
\title{GRAPE-5: A Special-Purpose Computer for $N$-body Simulation}

\author{
Atsushi {\sc Kawai},$^{1}$
Toshiyuki {\sc Fukushige},$^{1}$
Junichiro {\sc Makino},$^{2}$
and Makoto {\sc Taiji},$^{3}$
}
\affil{
$^1${
	Department of General Systems Studies,
	College of Arts and Sciences,
	University of Tokyo,
}\\
{
	3-8-1 Komaba,
	Meguro-ku,
	Tokyo 153
}\\
$^2${
Department of Astronomy, School of Science,
University of Tokyo,} \\
{\it
7-3-1, Hongo, Bunkyo-ku, Tokyo 113
}\\
$^3${
Institute of Statistical Mathematics,} \\
{
4-6-7 Azabu,
Minato-ku,
Tokyo 106
}
}

\authoremail{kawai@grape.c.u-tokyo.ac.jp}

\begin{abstract}
We have developed a special-purpose computer for gravitational
many-body simulations, GRAPE-5.  GRAPE-5 is the successor of
GRAPE-3. Both consist of eight custom pipeline chips (G5 chip and
GRAPE chip). The difference between GRAPE-5 and GRAPE-3 are: (1) The
G5 chip contains two pipelines operating at 80 MHz, while the GRAPE
chip had one at 20 MHz. Thus, the calculation speed of the G5 chip and
that of GRAPE-5 board are 8 times faster than that of GRAPE chip and
GRAPE-3 board. (2) The GRAPE-5 board adopted PCI bus as the interface
to the host computer instead of VME of GRAPE-3, resulting in the
communication speed one order of magnitude faster. (3) In addition to
the pure $1/r$ potential, the G5 chip can calculate forces with
arbitrary cutoff functions, so that it can be applied to Ewald or
P${}^3$M methods. (4)The pairwise force calculated on GRAPE-5 is about
10 times more accurate than that on GRAPE-3. On one GRAPE-5 board, one
timestep of 128k-body simulation with direct summation algorithm takes
14 seconds. With Barnes-Hut tree algorithm ($\theta = 0.75$), one
timestep of 10${}^6$-body simulation can be done in 16 seconds.
\end{abstract}

\section{Introduction}

In this paper, we describe the hardware architecture and performance
of GRAPE-5, the newest addition to GRAPE series of special-purpose
computers.

\begin{figure}[hbtp]
\centerline{\psfig{file=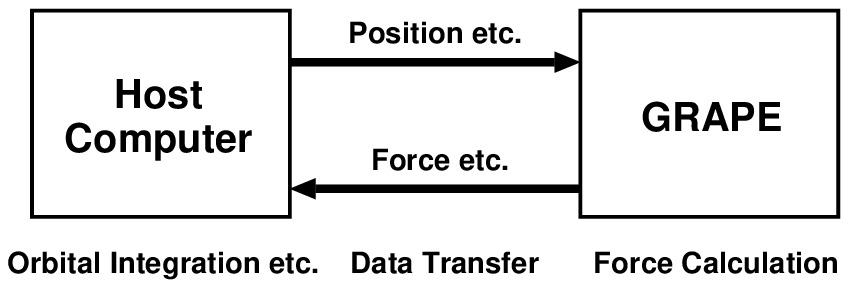,width=80mm}}
\caption{
Basic structure of GRAPE system.
}
\label{fig:grape}
\end{figure}

GRAPE (for ``GRAvity PipE''; see Sugimoto {\it et al.} 1990, Makino
and Taiji 1998) is a special-purpose computer to accelerate the
gravitational many-body simulation. It has pipelines specialized for
the force calculation, which is the most expensive part of the
gravitational many-body simulation.  All other calculations, such as
time integration of orbits, are performed on a host computer connected
to GRAPE.  Figure \ref{fig:grape} illustrates the basic structure of a
GRAPE system.  In the simplest case, the host computer sends positions
and masses of all particles to GRAPE.  Then GRAPE calculates the
forces between particles, and sends them back to the host computer.
The GRAPE system achieved high performance on the gravitational
many-body simulation through pipelined and highly parallelized
architecture specialized for the force calculation.

\begin{figure}[hbtp]
\centerline{\psfig{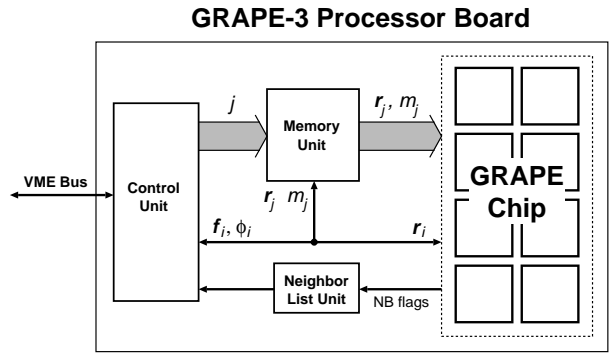}}
\caption{
Block diagram of the GRAPE-3 processor board.
}
\label{fig:g3b}
\end{figure}

GRAPE-3 (Okumura {\it et al.} 1993) is the first GRAPE system with
multiple force calculation pipelines. Figure \ref{fig:g3b} shows the
architecture of GRAPE-3A board with 8 pipelines. One pipeline is
integrated into the GRAPE chip. The peak performance of the GRAPE chip
is 0.6 Gflops at a clock cycle of 20 MHz, and that of a system which
contains 8 GRAPE chips is 4.8 Gflops.  Like GRAPE-1 (Ito {\it et al.} 
1990) and GRAPE-1A (Fukushige {\it et al.} 1991), GRAPE-3 uses the
number format with short mantissa, in order to reduce the chip
size. The r.m.s. relative error of the pairwise force is about 2\%,
which is low enough for most of collisionless simulations (Makino
1990, Hernquist {\it et al.} 1993, Athanassoula {\it et al.} 
1998). Copies of GRAPE-3 are used in many institute inside and outside
Japan ({\it e.g.} Brieu {\it et al.} 1995, Padmanabhan {\it et al.} 
1996, Steinmetz 1996, Nakasato {\it et al.}  1997, Klessen {\it et
al.} 1998, Mori {\it et al.} 1999, Theis and Spurzem 1999, Koda {\it
et al.} 1999).

GRAPE-5 is the successor of GRAPE-3. GRAPE-5 embodies the improvement
of a factor of 10 in calculation speed, communication speed, and
accuracy, over GRAPE-3. In addition, it can be applied to Ewald or
P${}^3$M algorithms, since it can evaluate the force and potential
with arbitrary cutoff.

The structure of this paper is as follows: In section
\ref{sec:algorithms} we briefly describe force calculation algorithms
used with GRAPE-5. In section \ref{sec:hardware} we describe the
hardware of the GRAPE-5 system. In section \ref{sec:accuracy} we
discuss the calculation accuracy of GRAPE-5. In section
\ref{sec:timing} we present the timing results. In section
\ref{sec:future} we discuss future prospects.

\section{Force Calculation Algorithms}\label{sec:algorithms}

In this section we briefly discuss force calculation algorithms used
on GRAPE-5. The $O(N^2)$ direct summation is the simplest algorithm.
To obtain the force on a particle, GRAPE-5 simply calculates and adds
up forces from all particles in the system. Though the direct
summation is simple and efficient for small-$N$ (say $N<$10${^5}$)
systems, for large-$N$ systems force calculation becomes expensive,
even with GRAPE hardware. In the following, we describe two algorithms
to reduce the calculation cost. First we describe the Barnes-Hut tree
algorithm (Barnes and Hut 1986), and then P${}^3$M (particle-particle
particle-mesh) method (Hockney and Eastwood 1981).

\subsection{Barnes-Hut Tree Algorithm}\label{sec:tree}

The Barnes-Hut tree algorithm (Barnes and Hut 1986) reduces the
calculation cost from $O(N^2)$ to $O(N\log N)$, by replacing forces
from distant particles by that from a particle at their center of mass
(or multipole expansion). Vectorization of tree algorithm is discussed
in Barnes (1990), Hernquist (1990) and Makino (1990). Parallelization
has been discussed in numerous articles (Salmon and Warren 1992,
Salmon {\it et al.} 1994, Dubinski 1996, Warren {\it et al.} 1997,
Yahagi {\it et al.} 1999).

The application of GRAPE hardwares to the tree algorithm are discussed
in Makino (1991) and Athanassoula {\it et al.}(1998). Vectorized
versions of tree algorithms based on Barnes' modified algorithm (1990)
are used in these articles. With this algorithm, we can use GRAPE more
efficiently than with the original algorithm. In the original
algorithm, tree traversal is performed for each particle. In the
modified algorithm, tree traversal is performed for a group of
neighboring particles and an {\it interaction list} is created. GRAPE
calculates the force from particles and nodes in this interaction list
to particles in the group.

The modified tree algorithm reduces the calculation cost of the host
computer by roughly a factor of $n_g$, where $n_g$ is the average
number of particles in groups.  On the other hand, the amount of work
on GRAPE increases as we increase $n_g$, since the interaction list
becomes longer.  There is, therefore, an optimal value of $n_g$ at
which the total computing time is minimum (Makino 1991). The optimal
value of $n_g$ depends on various factors, such as the relative speed
of GRAPE and its host computer, the opening parameter and number of
particles. For present GRAPE-5, $n_g=2000$ is close to optimal.

When we need high accuracy for the force calculation, we can use
P${}^2$M${}^2$ (pseudo-particle multipole method; Makino 1998). In the
algorithm described above, we can not handle higher-order terms of the
multipole expansion on GRAPE, since GRAPE can calculate $1/r^2$ force
only.  Thus, the amount of the calculation grows quickly when high
accuracy is required. In P${}^2$M${}^2$, high-order expansion terms
are expressed by forces from a small number of pseudo-particles, and
thus we can evaluate these terms on GRAPE. Using P${}^2$M${}^2$, Kawai
and Makino (1999) implemented arbitrary-order tree algorithm on
GRAPE-4. The timing results show that the calculation with
P${}^2$M${}^2$ is faster than that without P${}^2$M${}^2$, when the
total force error smaller than $\sim$ 10${}^{-4}$ is required.

\subsection{P${}^3$M Method}\label{sec:p3ma}

The P${}^3$M method (Hockney and Eastwood 1981) calculates
gravitational interaction under periodic boundary condition.  The
total force is divided into long-range and short-range forces. The
long-range (PM) force is computed in wave number space using fast
Fourier transform (FFT). The short-range (PP) force is calculated
directly.

Since the PM part takes care of the long-range interaction, the force
calculation in the PP part is not a pure $1/r^2$ gravity but with
cutoff. For example, Efstathiou and Eastwood (1981) used the following
form as the PP force exerted from a particle located at $\Vec{r}$
\begin{equation}
  \Vec{f(r')} = \frac{m (\Vec{r}-\Vec{r'})}{|\Vec{r}-\Vec{r'}|^3}g_{\rm P3M}(R),
\end{equation}
where $m$ is the mass of the particle and $\Vec{r'}$ is the position at
which the force is evaluated. The cutoff $g_{\rm P3M}$ is expressed
as
\begin{equation}\label{eq:gp3m}
\hspace*{-5mm}
g_{\rm P3M}(R) = \nonumber \\
\left\{
  \begin{array}{l}
    {\displaystyle
    1 - \frac{1}{140}(
                         224R^{3}-
                         224R^{5}+
                         70R^{6}
    } \nonumber \\[3mm]
    {\displaystyle
                         ~~~~+48R^{7}-
                         21R^{8}
                        )
    ~~~~{\rm for~~~~} 0 \le R < 1 } \nonumber \\[5mm]
    {\displaystyle
    1 - \frac{1}{140}(
                        12-
                        224R^{2}+
                        896R^{3}-
                        840R^{4}
    } \nonumber \\[3mm]
    {\displaystyle
                   ~~~~+224R^{5}+
                        70R^{6}-
                        48R^{7}+
                        7R^{8}
                        )
    } \nonumber \\[3mm]
    {\displaystyle
    ~~~~{\rm for~~~~} 1 \le R < 2 } \nonumber \\[5mm]
    {\displaystyle 0}
    ~~~~{\rm for~~~~} R \ge 2, \\
  \end{array}
  \right.
\end{equation}
where $R \equiv |\Vec{r}-\Vec{r_j}|/\eta$, and $\eta$ is a scale
length. If higher accuracy is desired, one can use a Gaussian cutoff
(Ewald 1921).  GRAPE-5 can calculate these forces using its
programmable cutoff table.

Brieu {\it et al.} (1995) implemented the P${}^3$M method on GRAPE-3.
Since GRAPE-3 can calculate force with Plummer softening only, they
expressed the PP force as a combination of three forces with different
Plummer softening radius. Therefore they used GRAPE-3 three times to
evaluate one PP force. This procedure is rather complex, and yet the
accuracy is rather low. GRAPE-5 evaluates this interaction with cutoff
in single call, and in high accuracy.

\section{Hardware of GRAPE-5 System}\label{sec:hardware}

In this section we describe the function and the architecture of
GRAPE-5 system. In section \ref{sec:g5arch} we show the overall
architecture of the GRAPE-5 system and overview the function of the
GRAPE-5 processor board. In section \ref{sec:g5chip} we give a
detailed description of the G5 chip, a custom chip for force
calculation. Sections \ref{sec:mu}--\ref{sec:iu} are devoted to
description of components of the processor board other than the G5
chip. Peak performance and other miscellaneous aspects of the board
are described in section \ref{sec:misc}.

\subsection{Overall Architecture}\label{sec:g5arch}

\begin{figure}[hbtp]
\centerline{\psfig{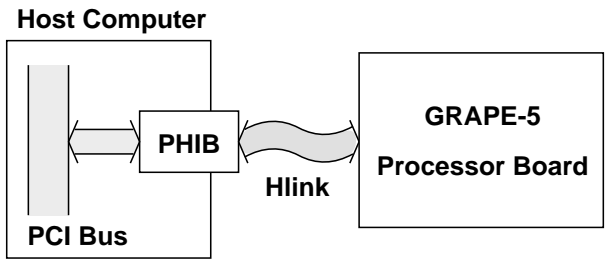}}
\caption{
Basic structure of the GRAPE-5 system.
}
\label{fig:g5sys}
\end{figure}

Figure \ref{fig:g5sys} shows the basic structure of the GRAPE-5
system.  It consists of three components: a GRAPE-5 processor board, a
PCI Host Interface Board (Kawai {\it et al.} 1997), and the host
computer.  The processor board is connected to PCI Host Interface
Board (hereafter we call PHIB) via Hlink (Makino {\it et al.} 1997).
PHIB is attached to PCI I/O bus of the host computer. PCI bus is an
I/O bus standard widely used from desktop PCs to supercomputers.

\begin{figure}[hbtp]
\centerline{\psfig{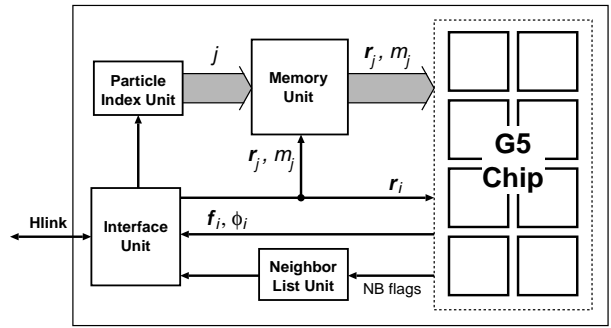}}
\caption{
Block diagram of the GRAPE-5 processor board.
}
\label{fig:g5b}
\end{figure}

Figure \ref{fig:g5b} shows the block diagram of the GRAPE-5 processor
board. It consists of five units: the G5 chips, the memory unit, the
particle index unit, the neighbor list unit, and the interface unit.
The G5 chip is a custom VLSI chip which integrates two pipeline
processors to calculate gravitational interactions.  The memory unit
supplies particle data to the G5 chips. The particle index unit
supplies indices of particles to the memory unit during force
calculation. This unit can supply indices in a special manner
optimized to the cell-index method to calculate a short-range force,
such as the PP force in P${}^3$M method. The neighbor list unit
constructs the list of the nearest neighbor particles. The interface
unit handles the communication with the host computer.
In the following subsections we describe these units.

\subsection{The G5 chip}\label{sec:g5chip}

\subsubsection{Function}

The G5 chip calculates the force exerted on particle $i$ at position
$\Vec{r}_i$. The force $\Vec{f}_i$ is expressed as
\begin{equation}
  \Vec{f}_{i} = \sum_j^N \Vec{f}_{ij},
\end{equation}
where $N$ is the number of particles and
\begin{equation}
\Vec{f}_{ij} =
  \left\{
  \begin{array}{ll}
    {\displaystyle
    \frac{m_j(\Vec{r}_j-\Vec{r}_i)}{r^3_{{\rm s,}ij}}
       g(r_{{\rm s,}ij}/\eta) } &
        {\rm for~~~} r_{{\rm s,}ij} \le r_{\rm cut}\\
    0 & {\rm for~~~} r_{{\rm s,}ij} > r_{\rm cut}.\\
  \end{array}
  \right.
  \label{eq:ppforce}
\end{equation}
Here $\Vec{f}_{ij}$ is the force (per unit mass) from particle $j$ to
particle $i$. Note that $\Vec{r}_j$ and $m_j$ are the position and the
mass of particle $j$, and that $r_{{\rm s,}ij}$ is the softened
distance between particle $i$ and $j$ defined as $r_{{\rm s,}ij}^2
\equiv |\Vec{r}_i - \Vec{r}_j|^2 + \epsilon_i^2$, where $\epsilon_i$
is the softening parameter for particle $i$. Here and hereafter, we
use index $i$ for the particle on which the force is evaluated and
index $j$ for particles that exert forces on particle $i$. The
function $g$ is an arbitrary cutoff function, $\eta$ is a scale length
for the cutoff function, and $r_{\rm cut}$ is the cutoff length.
The G5 chip also calculates potential $\phi_{i}$ associated with force
$\Vec{f}_i$, using cutoff function different from that for force
calculation.
During force calculation, The G5 chip asserts the neighbor flag if the
distance of particles $r_{{\rm s,}ij}$ is less than a given neighbor
radius, $h_{i}$.

\subsubsection{Overall architecture}

\begin{figure}[hbtp]
\centerline{\psfig{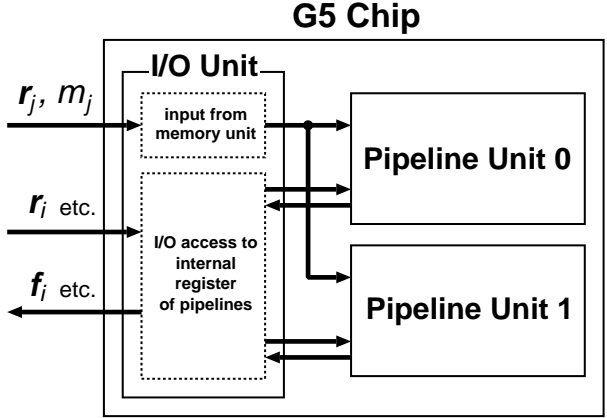}}
\caption{
Block diagram of the GRAPE-5 processor chip (G5 chip).
}
\label{fig:g5pipe1}
\end{figure}

The G5 chip consists of two pipeline units and one I/O unit, as shown
in figure \ref{fig:g5pipe1}. The pipeline unit calculates the
gravitational interaction. The I/O unit handles data transfer between
the pipeline unit and the external I/O port.

\begin{figure*}[hbtp]
\centerline{\psfig{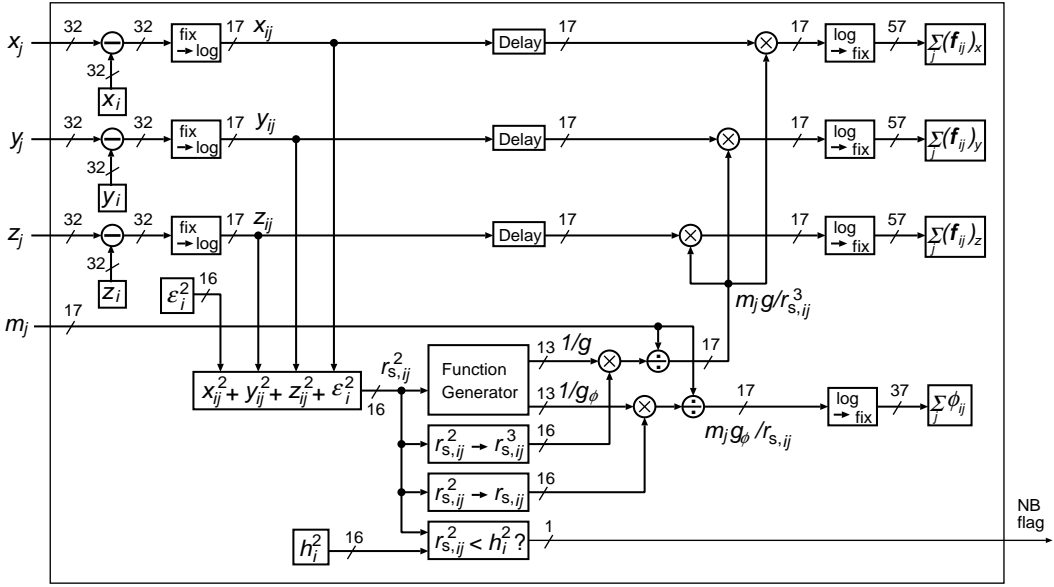}}
\caption{
Block diagram of the pipeline unit of the G5 chip.
}
\label{fig:g5pipe2}
\end{figure*}

\begin{figure}[hbtp]
\centerline{\psfig{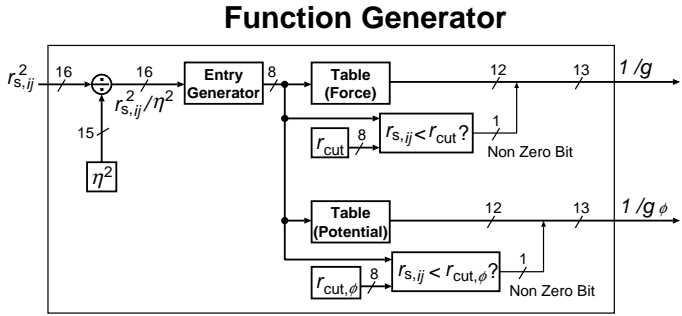}}
\caption{
Block diagram of the function generator of the G5 chip.
}
\label{fig:fev}
\end{figure}

Figure \ref{fig:g5pipe2} shows the block diagram of the pipeline unit.
The number attached to each line is the number of bits used. The
number format will be discussed later. The position vector $\Vec{r}_j$
and the mass data $m_j$ are supplied from the memory unit. The
position vector $\Vec{r}_i$, the softening parameter $\epsilon_i$, and
the neighbor radius $h_i$ are stored in the on-chip register before
the pipeline starts calculation. The pipeline unit calculates one
interaction per clock cycle, and accumulate it to on-chip registers.
The pipeline unit outputs the neighbor flag when $r_{{\rm s}, ij}^2 <
h_i^2$. The function generator calculates $1/g(r_{{\rm
s},ij}/\eta)$ and $1/g_{\phi}(r_{{\rm s},ij}/\eta)$ from
$r_{{\rm s},ij}^2$ and $\eta^2$.  Figure \ref{fig:fev} shows the block
diagram of the function generator.  The cutoff functions $g$ and
$g_{\phi}$ are evaluated by table lookup. The tables are implemented
as on-chip RAM blocks.


G5 chip has the {\it virtual multiple pipeline} architecture (Makino
{\it et al.} 1993) to reduce the necessary bandwidth of data transfer
during force calculation. In this architecture, one pipeline acts
as multiple pipelines operating at a slower speed.
The G5 chip has six virtual pipelines per real pipeline unit and
12 virtual pipelines in total.  Each real pipeline unit calculates the
forces exerted on 6 different particles.  Data for particle $j$ is
used for 6 clock cycles.  The memory unit operates at a clock
frequency 1/6 of that of the G5 chip. This architecture simplifies the
board design.

\begin{figure}[hbtp]
\centerline{\psfig{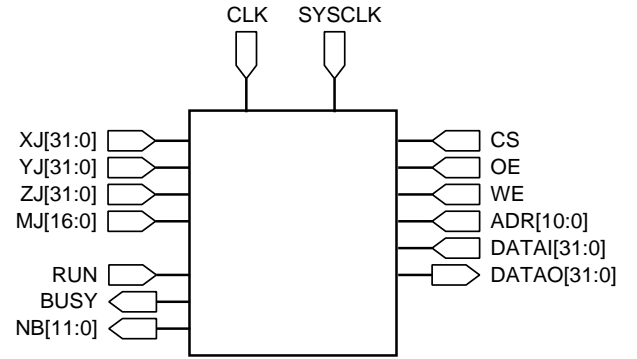}}
\caption{
I/O specification of the G5 chip.
}
\label{fig:g5io}
\end{figure}

Figure \ref{fig:g5io} shows the I/O specification of the G5 chip. It
has four input ports for particle data (XJ[31:0], YJ[31:0], ZJ[31:0],
and MJ[16:0]), one input and one output port to the host computer
(DATAI[31:0] and DATAO[31:0]), one address bus (ADR[10:0]), six
control input pins (CLK, SYSCLK, RUN, CS, OE and WE), and 13 output
pins (NB[11:0] and BUSY).
The CLK supplies clock signal for chip internal operations.  The
SYSCLK supplies clock signal for I/O operations. The frequency of
SYSCLK is 1/6 of that of CLK.

\subsubsection{Number format}\label{sec:num}

Following the design of GRAPE-1 (Ito {\it et al.} 1990) and GRAPE-3,
we adopted word length shorter than those used on conventional
computers for the G5 chip. The word length directly determines the
number of transistors. In order to achieve high performance at low
cost, it is crucial to use the minimum word length that ensures the
validity of the calculation. The word length used in the G5 chip are
shown in figure \ref{fig:g5pipe2} and figure \ref{fig:fev}. The number
of bits is attached to each data line.

We adopted the logarithmic format for most of the operations in the
pipeline unit except for the subtraction of the position vectors,
lookup of the cutoff table, and the final accumulation of the force.
We prefer the logarithmic format over the fixed-point format because
it has larger dynamic range for the same word length. Of course, the
usual floating-point format also can achieve a wider dynamic range. We
chose the logarithmic format because operations such as multiplication
and square root are easier to implement in the logarithmic format than
in the floating-point format. Although the addition becomes complex,
the logarithmic format is more efficient for the word length we used
for GRAPE-5.

In the logarithmic format, a positive, non-zero real number $x$ is
represented by its base-2 logarithm $y$ as
\begin{equation}\label{eq:log}
  x = 2^y.
\end{equation}
In G5 chip, we use 15 bits to express $y$ in unsigned fixed-point
format, with 8 bits below binary point. The 16th bit indicates whether
$x$ is 0 or not (non-zero bit), and the 17th bit indicates the
sign. Thus the total number of bits per word is 17.  This format can
express real number in the range of $[1,2^{128})$, and its resolution
is $2^{1/256}-1 \simeq 0.003$.

We use 32-bit fixed-point 2's-complement format for each component of
the position vector $\Vec{r}_i$ and $\Vec{r}_j$, in order to guarantee
uniform resolution and to simplify implementation of the periodic
boundary condition. The selection of the minimum image is performed
automatically, by setting the size of the box length to the maximum
expressible number ($2^{32}-1$). This format gives a spatial
resolution of $2^{-32} \simeq 10^{-9}$.

We use 64-bit and 50-bit fixed-point format for accumulation of the
force and potential, respectively. We adopt the fixed-point format
because the adder (accumulator) is simpler and its cost is lower in
this format than that in logarithmic or floating-point format.

For conversion between the fixed-point format and the logarithmic
format, we use format converters described in section \ref{sec:f2l}
and \ref{sec:l2f}. For addition in the logarithmic format, we use
logarithmic adder described in section \ref{sec:add}.

\subsubsection{Format converter (fixed point to logarithmic)}\label{sec:f2l}

Here we describe the hardware to convert the fixed-point format to the
logarithmic format. Output number has $\beta = 2+\gamma+\delta$ bits,
where $\gamma$ and $\delta$ are the number of bits above and below the
binary point, respectively. In the case of the G5 chip, $\beta = 17$,
$\gamma = 7$ and $\delta = 8$.

\begin{figure}[hbtp]
\centerline{\psfig{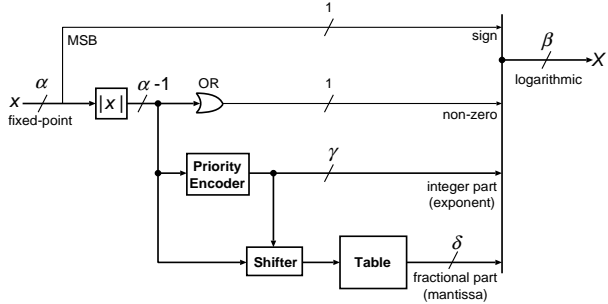}}
\caption{
Block diagram of the format converter (fixed-point format to
logarithmic format).
}
\label{fig:f2l}
\end{figure}

Figure \ref{fig:f2l} shows the block diagram of the format converter.
First we convert the 2's complement format to the sign+magnitude
format. Then we calculate the ``integer'' part of logarithm (upper
$\gamma$ bits) using a priority encoder.

The fractional part of logarithm is calculated as follows.  We first
normalize the magnitude of $x$ by removing leading zeros. This is done
by a logical shifter with control input from the priority encoder. The
output of the shifter is then supplied to a table which convert a
normalized number to the base-2 logarithm. In case of G5 chip, the
table has 512 ($= 2^9$) entries to ensure that the round-off error
generated at conversion is small.

\subsubsection{Logarithmic adder}\label{sec:add}

The (unsigned-)logarithmic adder performs addition of two positive
number $X$ and $Y$ in logarithmic format. The design of logarithmic
adder of G5 chip is basically the same as that of GRAPE chip. We
designed it using the following relation
\begin{eqnarray}
  Z & \equiv & \log_2(2^X+2^Y) \nonumber \\
    & = & \log_2 2^X + \log_2(1+2^Y/2^X) \nonumber \\
    & = & X + \log_2(1+2^{Y-X})
\end{eqnarray}
for $X \ge Y$, or,
\begin{equation}\label{eq:add}
  Z = {\rm max}(X, Y) + \log_2(1+2^{-|X-Y|})
\end{equation}
for general $X$ and $Y$ (Kingsbury and Rayner 1971, Swartzlander and
Alexopoulos 1975).

\begin{figure}[hbtp]
\centerline{\psfig{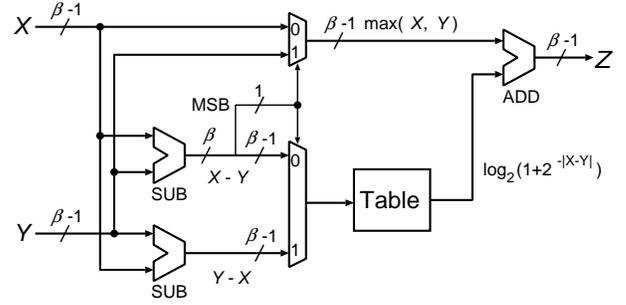}}
\caption{
Block diagram of the logarithmic adder.
}
\label{fig:add}
\end{figure}

Figure \ref{fig:add} shows the block diagram of the logarithmic adder. 
First we calculate $X-Y$ and $Y-X$, and choose the positive one using
a multiplexer. Then we use table lookup to obtain
$\log_2(1+2^{-|X-Y|})$ from $|X-Y|$. Finally we obtain $Z$ by adding
the output of the table to the larger one of $X$ and $Y$.

In practice, we do not need to prepare table for all possible values
of $|X-Y|$. If $\log_2(1+2^{-|X-Y|}) < 2^{1/2^\delta} - 1$, the result
of addition $Z$ is equal to $X$ (assuming $X > Y$), after we properly
rounded the result. Thus, in the case of the G5 chip with $\delta =
8$, we need table only for $|X-Y| < - \log_2[2^{(2^{2^{-8}}-1)}-1]
\simeq 9.1$ (see Makino and Taiji 1998, section 4.6 for detailed
discussion).

\subsubsection{Cutoff function generator}\label{sec:fev}

Figure \ref{fig:fev} shows the block diagram of the cutoff function
generator. The cutoff function $g(r_{{\rm s},ij}/\eta)$ and
$g_{\phi}(r_{{\rm s},ij}/\eta)$ are calculated from $r_{{\rm s},ij}^2$
and $\eta^2$. First we divide $r_{{\rm s},ij}^2$ by $\eta^2$ using
subtracter and input the result to the entry generator. Then we supply
the output of the entry generator to the cutoff function tables, and
obtain $1/g$ and $1/g_{\phi}$ as output of those tables.  The contents
of the cutoff function tables are set by the user before the force
calculation starts.

The cutoff function table is realized by an on-chip RAM. The RAM table
consumes significantly larger number of transistors per bit than ROM
tables used in the format converter and the logarithmic adder. In
order to achieve a good cost performance, it is important to keep the
size of the table as small as possible.

\begin{figure}[hbtp]
\centerline{\psfig{file=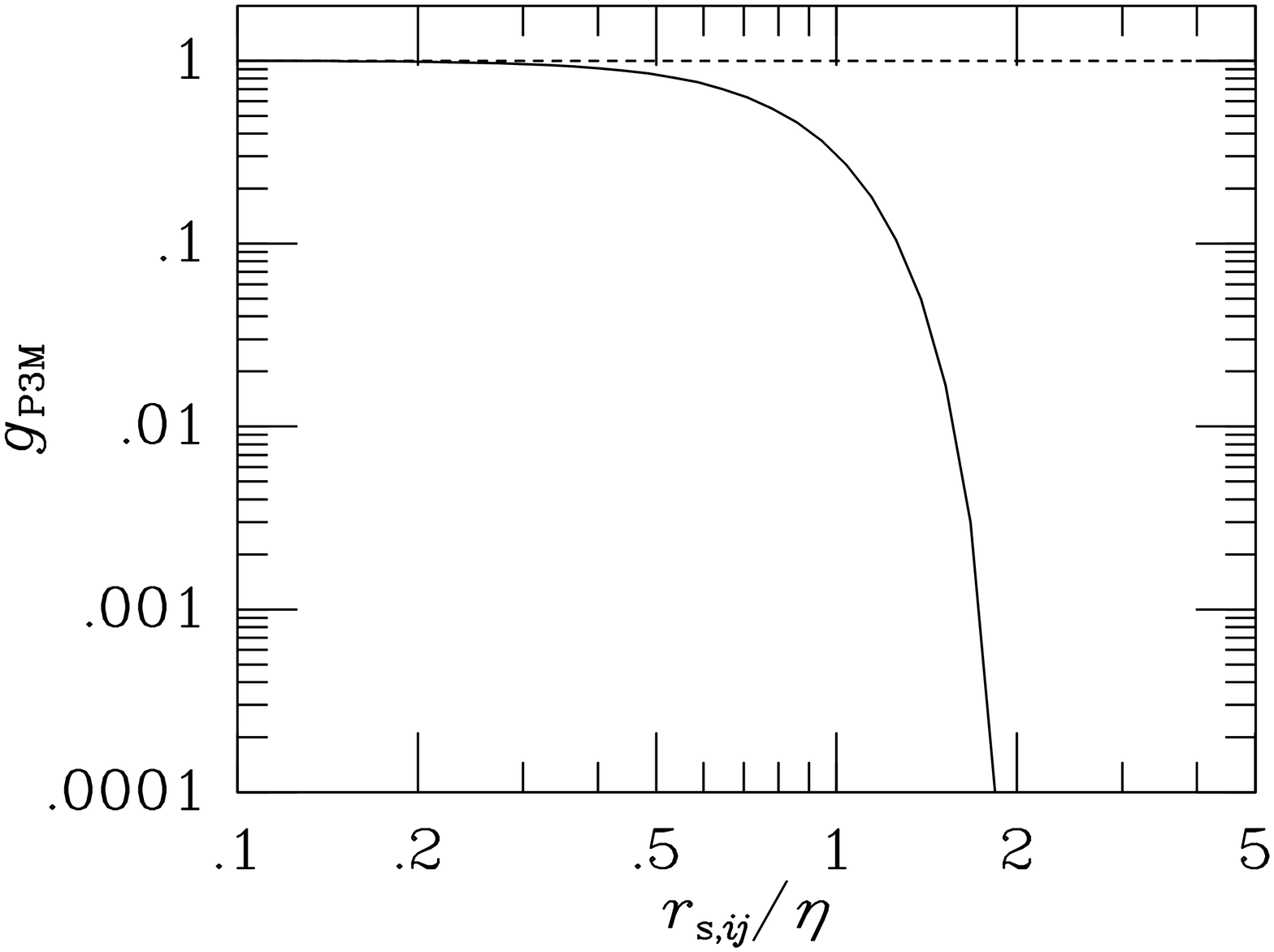,width=80mm}}
\caption{
The cutoff function for P$^3$M method, $g_{\rm P3M}$, given by
equation (\protect{\ref{eq:gp3m}}).
}
\label{fig:funcg}
\end{figure}

We reduce the size of the table by taking advantage of the nature of
the cutoff function. The cutoff function $g$ converges to 1 when
$r_{{\rm s},ij}/\eta$ approaches to 0, and converges to 0 when
$r_{{\rm s},ij}/\eta \gg 1$ (see figure \ref{fig:funcg}). Therefore we
need high resolution only when $r_{{\rm s},ij}/\eta \sim 1$. Using
these characteristics, we can reduce the number of entries to the
cutoff table for small $r_{{\rm s},ij}/\eta$ ($\lesssim 0.1$) and
large $r_{{\rm s},ij}/\eta$ ($\gtrsim 2$).

The entry generator reduces the number of entries to the cutoff
function table in the following two steps. At the first step, the
input $r_{{\rm s},ij}^2/\eta^2$ expressed in 15-bit logarithmic format
(without sign and non-zero bit) is converted to a 9-bit integer number.
The conversion is expressed as
\begin{equation}
  B = 128 \cdot 2^{A/512},
\end{equation}
where $A$ is the logarithmic part of the input number interpreted as
unsigned integer, and $B$ is the output. This conversion reduces the
number of entries at small $r_{{\rm s},ij}/\eta$.
At the second step, we reduce the entries at large $r_{{\rm
s},ij}/\eta$ using the conversion given in table \ref{tab:funcg}. An
8-bit integer number is obtained as the conversion result.  The
maximum $r_{{\rm s},ij}/\eta$ that is expressible in the final format
is $3.0$.  This value is large enough for typical cutoffs such as
$g_{\rm P3M}$ given by equation (\ref{eq:gp3m}) and a Gaussian cutoff,
since the value of cutoff at $r_{{\rm s},ij}/\eta = 3.0$ is smaller
than the force calculation error. In the actual implementation, the
entry generator directly converts the input to the final format by
single table lookup.

\begin{table}[hbtp]
\caption{Entry reduction procedure for cutoff function
$g$}\label{tab:funcg}
\begin{center}
\begin{tabular}{lll}
\hline
\hline
$r_{{\rm s},ij}/\eta$ & input            & output\\
                      & ($B$ in decimal) & (in binary)\\
\hline
0.0 - 1.5 & ~~~0 - 191 & $B[7]$~~~~~~~~-~~~~~~$B[0]$ \\
1.5 - 2.0 & 192 - 255 & 1~~1~~0~~$B[5]$~~-~~$B[1]$ \\
2.0 - 3.0 & 256 - 383 & 1~~1~~1~~$B[6]$~~-~~$B[2]$ \\
3.0 -     & 384 -     & 1~~1~~1~~1~~1~~1~~1~~1 \\
\hline
\end{tabular}
\end{center}
\end{table}

In order to reduce the size of the RAM table, we should reduce not
only the number of entries but also the word length. The logarithmic
format we used has 17 bits. We do not need sign and non-zero bits for
$g$. In addition, we do not need the full 7-bit integer part, since
$g$ smaller than $1/256$ can be treated as zero without affecting the
overall accuracy. So the minimum number of bits above binary point is
three. We choose to assign 4 bits above the binary point. The table
actually stores $1/g$ instead of $g$, since the format can only
express numbers not smaller than 1.

\subsubsection{Format converter (logarithmic to fixed point)}\label{sec:l2f}

\begin{figure}[hbtp]
\centerline{\psfig{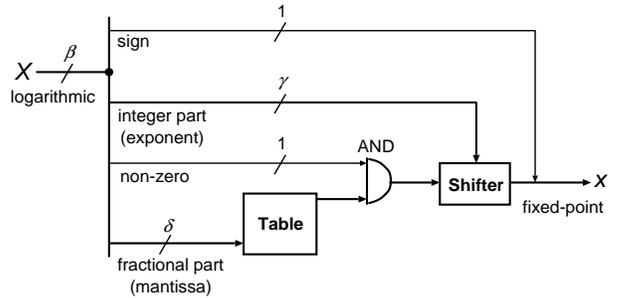}}
\caption{
Block diagram of the format converter (logarithmic format to
fixed-point format).
}
\label{fig:l2f}
\end{figure}

Figure \ref{fig:l2f} shows the block diagram of the circuit to convert
the logarithmic format to the fixed-point format. First we convert the
fractional part of the logarithmic format to a normalized number
(exponential of the input) by table lookup.  Then we shift it
according to the integer part of the logarithmic format.

\subsubsection{Packaging and other miscellaneous aspects}

The G5 chip is fabricated by NEC Corporation using 0.5 $\mu$m gate
array process (CMOS-8L family).  It consists of $\sim$ 200 K gates
(nominal 300 K gates with 65 \% usage) and is packaged in a ceramic
pin-grid array with 364 pins. It operates at 80MHz clock cycle. The
power supply voltage is 3.3 V and the power consumption is about 10 W. 
We designed the G5 chip using logic synthesis tool (AutoLogic II by
Mentor Graphics Corp.). All design is expressed in the VHDL language.

\subsection{Memory Unit}\label{sec:mu}

The memory unit supplies position vectors $\Vec{r}_j$ and masses $m_j$
of the particles to the G5 chip according to the indices supplied by
the particle index unit. The particle data $\Vec{r}_j$ and $m_j$ are
shared by all G5 chips.
The memory unit is composed of four 4 Mbit (128 Kword $\times$ 32 bit)
synchronous SRAM chips. Three are for position vectors and one is for
masses. This unit can store up to 131072 particles. To simulate a
system with more than 131072 particles, we need to divide the
particles into subgroups each of which includes less than 131072
particles. We can calculate the total forces by summing up the partial
forces from each subgroup.

\subsection{Particle Index Unit}\label{sec:piu}

The particle index unit supplies particle indices (memory address) to
the memory unit. This unit is optimized for the cell-index method
(also known as the linked-list method, Quentrec and Brot 1975, Hockney
and Eastwood 1981). We adopt the hardware design used in MD-GRAPE
(Fukushige {\it et al.} 1996).

The cell-index method is a scheme to reduce the calculation cost of
the short-range force. In this method, the cube which includes entire
simulation space (simulation cube) is divided into $M^3$ cells where
$M$ is the number of cells in one dimension. Usually, we set $M$ to
the largest integer not exceeding $L/r_{\rm cut}$, where $r_{\rm cut}$
is the cutoff length of the short-range force and $L$ is the side
length of the simulation cube. In order to calculate the forces on
particles in a cell, we need to calculate the contributions only from
the particles in 27 neighbor cells. Therefore, the calculation cost is
reduced by a factor of 27/$M^3$. One could also further reduce the
calculation cost by reducing the cell size and calculating the
contribution from all cells whose distance from the cell in question
is not exceeding $r_{\rm cut}$.

\begin{figure}[hbtp]
\centerline{\psfig{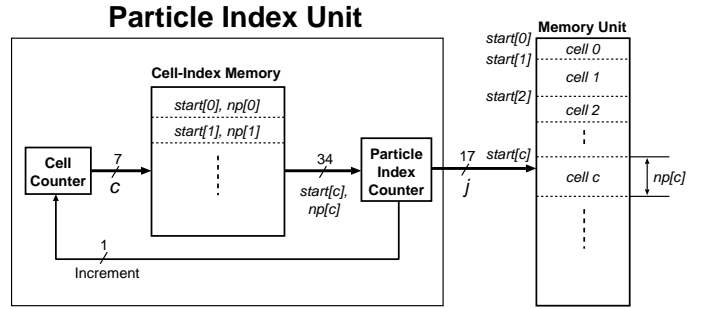}}
\caption{
Block diagram of the particle index unit of the GRAPE-5 processor board.
}
\label{fig:piu}
\end{figure}

Figure \ref{fig:piu} shows the block diagram of the particle index
unit. It consists of the cell-index memory and two counters: the cell
counter and the particle index counter. When we store particles to the
memory unit, we rearrange the order of particles so that particles in
the same cell occupy consecutive locations. Thus, we can specify all
particles in one cell by its start and end addresses.

The cell counter is a 7-bit counter, and the particle index counter is
a 17-bit counter. These two counters and cell-index memory are
packaged in an EPF10K30 FPGA device (Field Programmable Gate
Array, Altera Corp.), together with the interface unit.

\subsection{Neighbor List Unit}\label{sec:nlu}

\begin{figure}[hbtp]
\centerline{\psfig{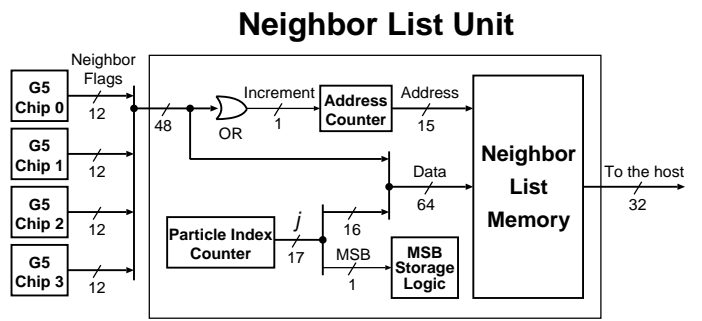}}
\caption{
Block diagram of the neighbor list unit of the GRAPE-5 processor board.
}
\label{fig:nb}
\end{figure}

The neighbor list unit stores the list of the nearest neighbors for
particle $i$.  Particle $j$ is the neighbor of particle $i$ if
$r_{{\rm s,}ij} < h_i$.  One neighbor list unit handles the neighbors
of particles calculated on four G5 chips. So we have two neighbor list
units on one board. Figure \ref{fig:nb} shows the block diagram of the
neighbor list unit.  Each of four G5 chips outputs neighbor flags for
12 particles at every clock cycle of the board. Thus the neighbor list
unit receives the flags for 48 particles. These neighbor flags are
stored together with the corresponding particle index $j$ in the
neighbor list memory if any of them is asserted. The host computer
reads the data from the neighbor list memory after the force
calculation is finished.

The neighbor list memory is composed of two 1 Mbit (32 Kword $\times$
32 bit) synchronous SRAMs. It can hold up to 32768 neighbors for 48
particles. The memory stores 48 flag bits and lower 16 bits of
particle index. The MSB of the particle index is stored in the FPGA
chip which implements other functions of neighbor list as well. The
MSB of the particle index changes from zero to one only once, since
the particle index always increases. Therefore, we need to store only
the value of the internal address counter of the neighbor list at
which MSB of the index counter is first set. Using the value of this
register, we can recover the MSB of particle index on the host
computer.

The logic for the neighbor list unit other than the neighbor list
memory is implemented in an EPF10K30 FPGA device.

\subsection{Interface Unit}\label{sec:iu}

The interface unit controls communication between the host computer
and the GRAPE-5 system. It recognizes the following five commands: (1)
receive data for particle $j$; (2) receive data for particle $i$; (3)
start the force calculation; (4) send back the calculated force
$\Vec{f}_i$ and potential $\phi_i$; (5) send back the neighbor list.
The control logic is implemented in an EPF10K30 FPGA device,
together with the particle index unit.

\subsection{Peak Performance and Other Miscellaneous Aspects}\label{sec:misc}

The peak performance of a GRAPE-5 board is 38.4 Gflops. The G5 chip
calculates 1.6 $\times$ 10$^8$ interactions per second. It calculates
two pairwise interactions in each clock cycle, and operates at a clock
cycle of 80 MHz. If we count the calculation of the gravitational
force as 30 floating-point operations, the peak performance of the G5
chip is equivalent to 4.8 Gflops. Thus the peak speed of a board with
eight G5 chips is 38.4 Gflops. The sustained performance is discussed
in section \ref{sec:timing}.

\begin{figure*}[hbtp]
\centerline{\psfig{file=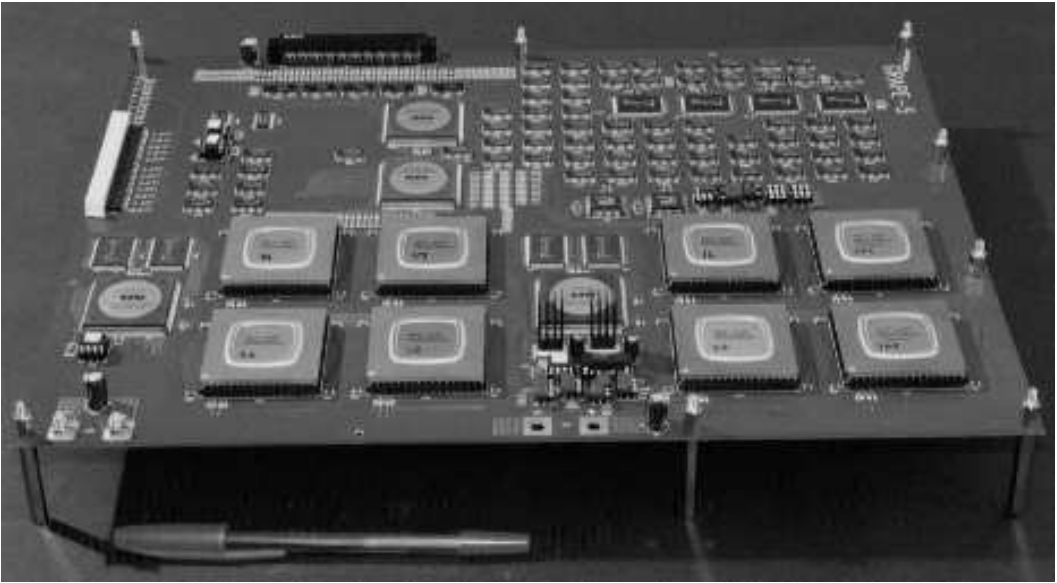,width=140mm}}
\caption{
Photograph of the GRAPE-5 processor board.
}
\label{fig:g5photo}
\end{figure*}

Figure \ref{fig:g5photo} shows photograph of the GRAPE-5 processor
board. The size of the board is 275 mm $\times$ 367 mm. Power supply
voltage is 3.3 V and the total power consumption is about 70 W.

We started designing the G5 chip in 1996 May. The chip was completed
in 1998 June. The first prototype of GRAPE-5 board with four G5 chips
was completed in 1998 October. The production version of the processor
board was completed in 1999 April.

\section{Accuracy of Force Calculation}\label{sec:accuracy}

In this section we discuss the calculation accuracy of the force from
one particle. In section \ref{sec:accpure}, we discuss the accuracy of
the $1/r^2$ gravity with softening. In section \ref{sec:acccutoff}, we
discuss the accuracy of the force with cutoff.

\subsection{Accuracy of $1/r^2$ Force}\label{sec:accpure}

A detailed analysis of the error propagation process (Makino {\it et
al.} 1990) gives an estimate of the relative error of the force from
one particle as
\begin{eqnarray}
  S_r(\Vec{f}) & \equiv &
                 \frac{\left<|\hat{\Vec{f}}-\Vec{f}|^2\right>}{f^2}
                 \nonumber \\
               & \lesssim &
                 \left(
                     \frac{18r^2}{r_{{\rm s},ij}^4} -
                     \frac{12}{r_{{\rm s},ij}^2} +
                     \frac{6}{r^2}
                 \right)\epsilon_i^2 \\ 
&&               + \left(
                     \frac{9r^4}{r_{{\rm s},ij}^4} -
                     \frac{6r^2}{r_{{\rm s},ij}^2} +
                     \frac{15}{2}
                 \right)\epsilon_f^2.
\end{eqnarray}
If $\epsilon \ll r$, the inequality becomes
\begin{eqnarray}
  S_r(\Vec{f}) & \lesssim &
                 \frac{12}{r^2}\epsilon_i^2+\frac{21}{2}\epsilon_f^2 \label{eq:pperr}.
\end{eqnarray}
Here $\Vec{f}$ and $\hat{\Vec{f}}$ are exact and calculated forces
from a particle at distance $r$. The parameter $\epsilon_i$ is the
r.m.s. absolute error of the fixed-point format used for the position
vectors ($\Vec{r}_i$ and $\Vec{r}_j$), and $\epsilon_f$ is the
r.m.s. relative error of the logarithmic format used for internal
number expression.
The first term of the right hand side of inequality (\ref{eq:pperr}) is
due to the round-off error of $\Vec{r}_i$ and $\Vec{r}_j$ expressed in
fixed-point format. The second term is due to the error of internal
calculation in logarithmic format.

As we described in section \ref{sec:num}, we used 32-bit fixed-point
format for position data and 17-bit (8 bits for fractional part)
logarithmic format for internal number expression. Therefore,
$\epsilon_i$ and $\epsilon_f$ are expressed as
\begin{equation}\label{eq:ei}
  \epsilon_i = \frac{r_{\rm max}}{2^{33} \sqrt{3}}
             \simeq 6.7 \times 10^{-11} r_{\rm max}
\end{equation}
and
\begin{equation}\label{eq:ef}
  \epsilon_f = \frac{\ln{2}}{512 \sqrt{3}}
             \simeq 7.8 \times 10^{-4}.
\end{equation}
Here $r_{\rm max}$ is the maximum value of the coordinates. Note that
$r_{\rm max}$ can be specified by software.

\begin{figure}[hbtp]
\centerline{\psfig{file=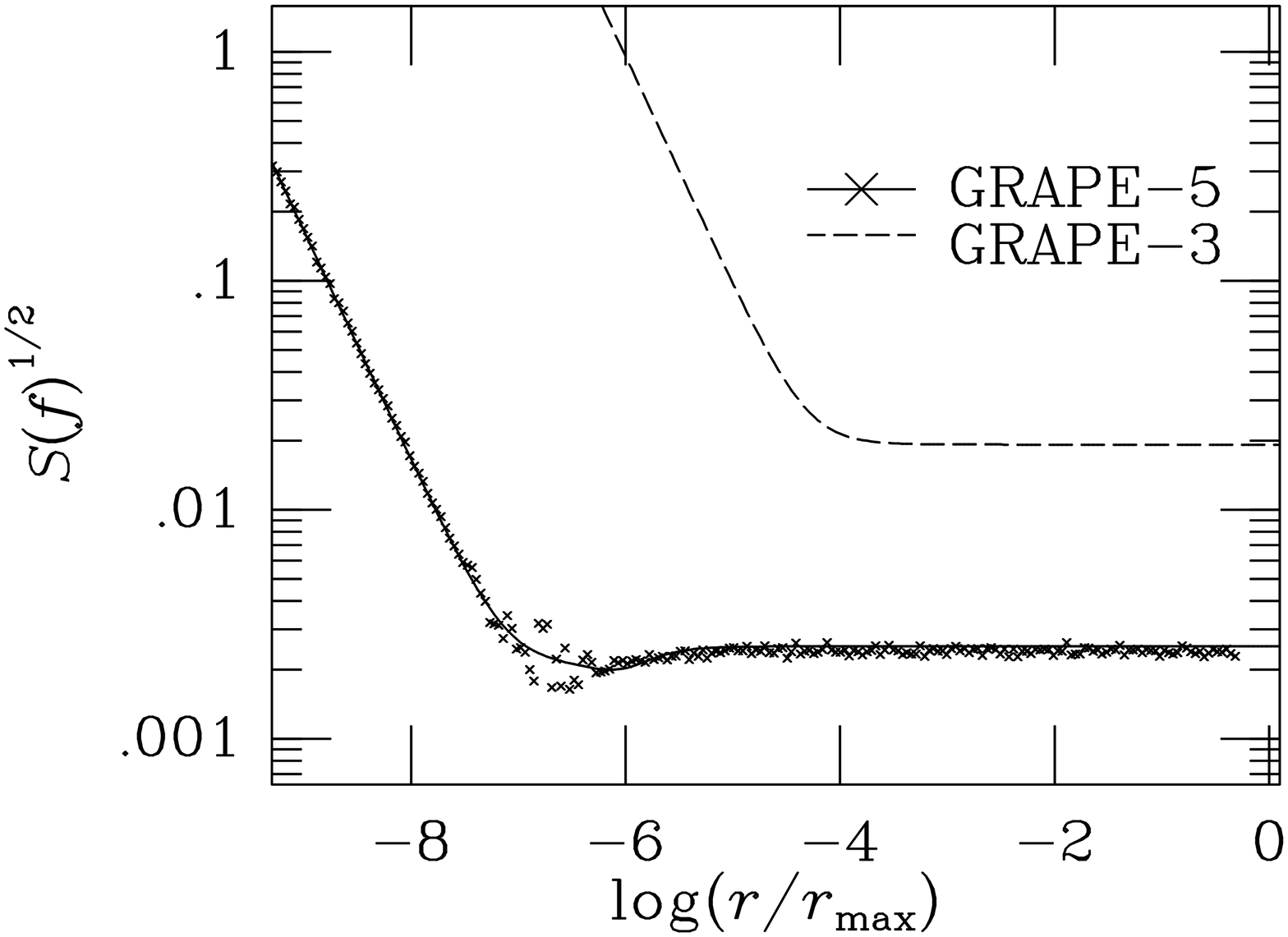,width=80mm}}
\caption{
Estimated error of the force from one particle in the case of pure
$1/r^2$ force with softening, as functions of distance $r$ scaled with
$r_{\rm max}$. The solid and dashed curves are theoretical estimate
for GRAPE-5 and GRAPE-3, respectively.  The crosses are errors
obtained with GRAPE-5.  The softening parameter is $\epsilon = 10^{-6}
r_{\rm max}$.  Values are averaged over 1000 trials.
}
\label{fig:err53}
\end{figure}

Figure \ref{fig:err53} shows the theoretical and measured values of
the error $S_r(\Vec{f})^{1/2}$ as functions of $r$.  The ``exact''
force $\Vec{f}$ is calculated using IEEE-754 standard 64-bit
arithmetic, and the force $\hat\Vec{f}$ is obtained from the software
emulator that gives the same results as G5 chip at bit level. The
softening parameter is set to $\epsilon = 10^{-6} r_{\rm max}$. The
mass is set to $m = m_{\rm min}$, where $m_{\rm min}$ is the smallest
mass with which the force from a particle located at far most distance
($r_{\rm max}$) does not underflow.
The theoretical estimate and measured errors show a good
agreement. For $r$ in the range of $[10^{-6} r_{\rm max}, r_{\rm
max}]$, the relative force error is around 0.3\%.  We also plot
theoretical estimate of error of GRAPE-3 in Figure \ref{fig:err53}.
We can see that the G5 chip has 10 times higher accuracy and the
dynamic range $10^3$ times wider than those of the GRAPE Chip.

\begin{figure}[hbtp]
\centerline{\psfig{file=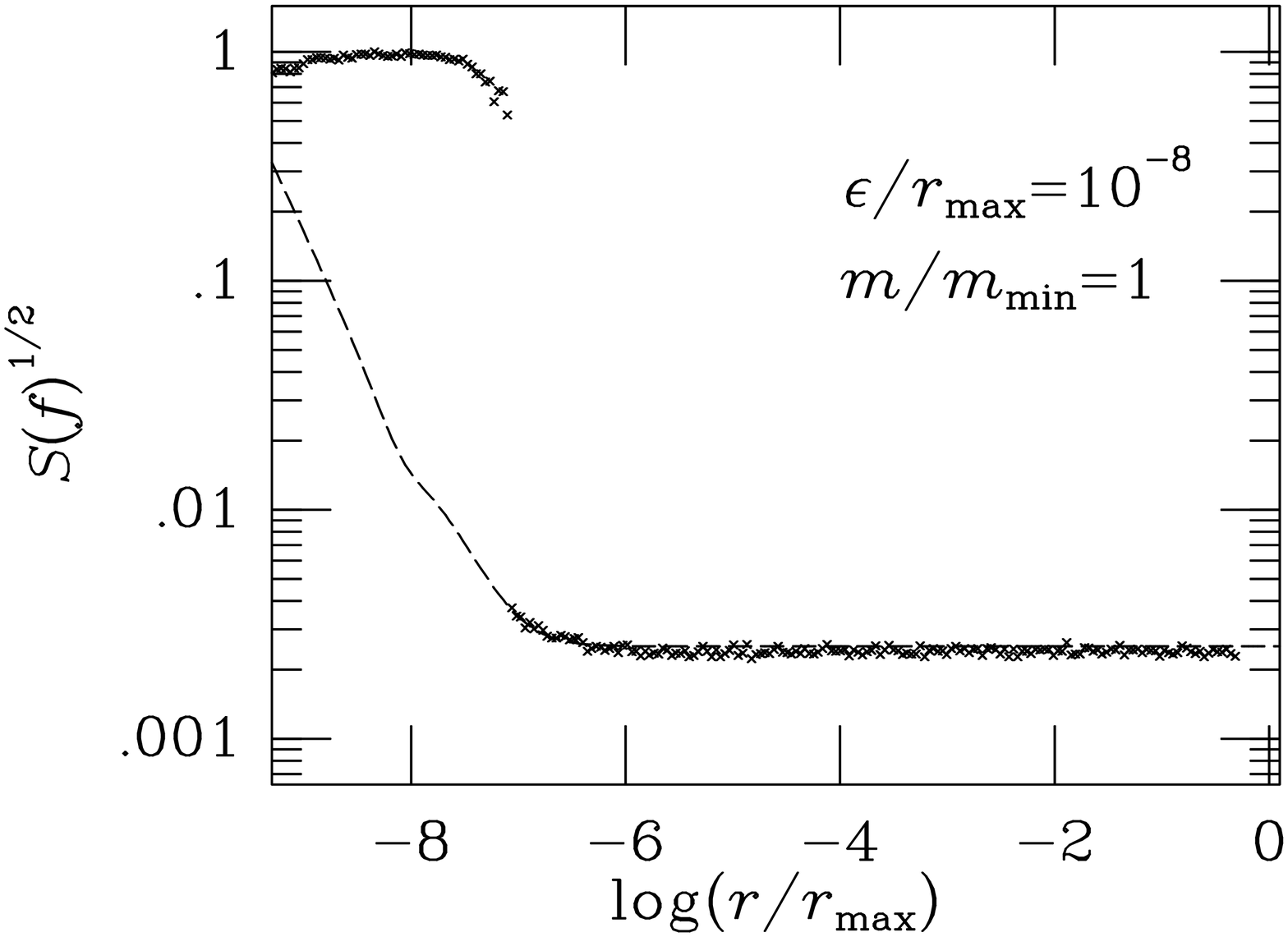,width=80mm}}
\caption{
Same as figure \protect{\ref{fig:err53}} but for smaller
softening. Crosses are measured errors for softening parameter
$\epsilon = 10^{-8} r_{\rm max}$ and mass $m = m_{\rm min}$. The
dashed curve is the theoretical estimate.
}
\label{fig:err5ovflw}
\end{figure}

In the case of calculation with an extremely small softening, the
force from close particle can overflow, {\it i.e.}, the force can be
too large to be expressed in the format we adopted for force
representation. Since the maximum force from one particle scales as
$m/\epsilon^2$, the minimum softening parameter $\epsilon_0(m)$ with
which the force does not overflow is expressed as
\begin{equation}\label{eq:ovflw}
\epsilon_0(m) = \left(\frac{m}{m_{\rm min}}\right)^{1/2}\epsilon_{\rm min},
\end{equation}
where $\epsilon_{\rm min} \simeq 10^{-7} r_{\rm max}$. Figure
\ref{fig:err5ovflw} shows such an extreme case. The force error for a
small softening($\epsilon = 10^{-8} r_{\rm max} \simeq 10^{-1}
\epsilon_{\rm min}$) is plotted. We can see that the force overflows
at $r \simeq 10^{-7} r_{\rm max}$.

\begin{figure}[hbtp]
\centerline{\psfig{file=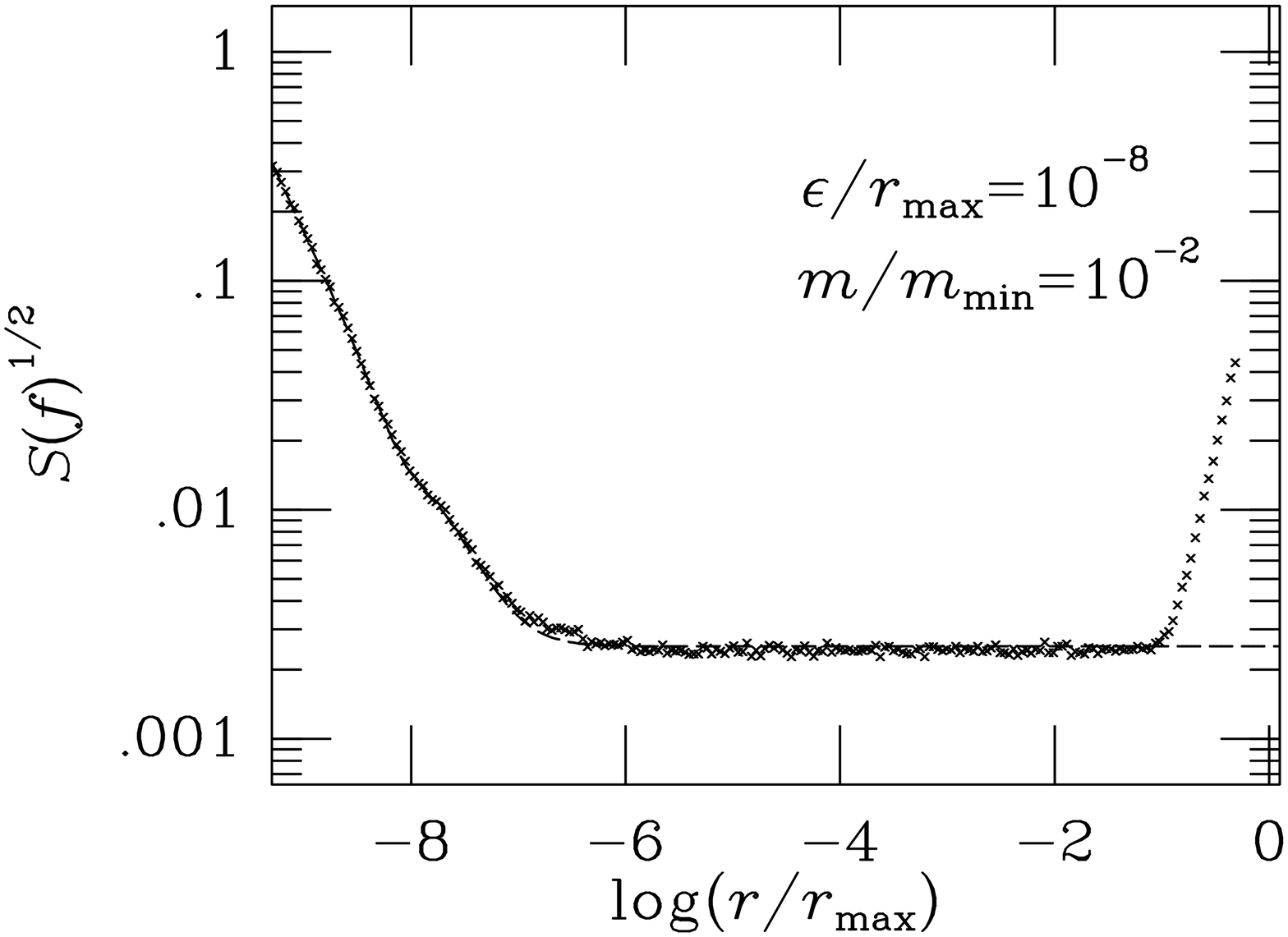,width=80mm}}
\caption{
Same as figure \protect{\ref{fig:err5ovflw}} but for smaller mass $m =
10^{-2} m_{\rm min}$.
}
\label{fig:err5udflw}
\end{figure}

We can avoid overflow of the force using masses smaller than $m_{\rm
min}$. Equation (\ref{eq:ovflw}) implies that $\epsilon_0(m)$ is
smaller than $\epsilon_{\rm min}$, if $m$ is smaller than $m_{\rm
min}$. In this case, however, the force underflows at large $r$, {\it
i.e.}, the force from distant particle is so small that a part of its
lower bits are lost. The maximum distance $r_0(m)$ with which the
force does not underflow is expressed as
\begin{equation}\label{eq:udflw}
r_0(m) = \left(\frac{m}{m_{\rm min}}\right)^{1/2} r_{\rm max}.
\end{equation}
Figure \ref{fig:err5udflw} shows the force error for softening smaller
than $\epsilon_{\rm min}$ ($\epsilon = 10^{-8} r_{\rm max} \simeq
10^{-1} \epsilon_{\rm min}$) and mass smaller than $m_{\rm min}$ ($m =
10^{-2} m_{\rm min}$). We can see that the force does not overflow
even with softening smaller than $\epsilon_{\rm min}$, although it
underflows at $r > 10^{-1} r_{\rm max}$.
This technique to use masses smaller than $m_{\rm min}$ is
particularly useful with the tree algorithm. In the tree algorithm,
distant nodes typically represent many particles, while close nodes
usually represent fewer number of particles. Thus, both the overflow
and underflow are automatically avoided.

\subsection{Accuracy of Force with Cutoff}\label{sec:acccutoff}

First we estimate the accuracy of the cutoff function table.  The
error of the table output is estimated as
\begin{equation}
|\hat{g}-g| \le g' \zeta_i + g \zeta_f \label{eq:errcutoff},
\end{equation}
where $g$ is the exact value of arbitrary cutoff function, $\hat{g}$
is the output of the table, and $g'$ is the derivative of $g$. The
parameter $\zeta_i$ is the absolute error of the table entry for the
worst case, and $\zeta_f$ is the relative error of the table output
for the worst case. The first term of the right hand side of
inequality (\ref{eq:errcutoff}) is due to the round-off error of the
table entry. The second term is due to the round-off error of the
table output.

The resolution of the entry of the table depends on the magnitude of
the entry itself, as shown in table \ref{tab:funcg}. The table entry
$r_{{\rm s},ij}/\eta$ has the finest resolution ($1/128$) in the range
of [0.0, 1.5], and the resolution doubles at $r_{{\rm s},ij}/\eta =
1.5$ and $r_{{\rm s},ij}/\eta = 2.0$.  Therefore $\zeta_i$ is given by
\begin{equation}
  \zeta_i = \frac{p}{256},
\end{equation}
where
\begin{equation}
  p = \left\{
    \begin{array}{ll}
           1 & ~~~{\rm for}~~~0.0 < r_{{\rm s},ij}/\eta < 1.5 \\
           2 & ~~~{\rm for}~~~1.5 < r_{{\rm s},ij}/\eta < 2.0 \\
           4 & ~~~{\rm for}~~~2.0 < r_{{\rm s},ij}/\eta < 3.0 \\
      \infty & ~~~{\rm for}~~~3.0 < r_{{\rm s},ij}/\eta .
    \end{array}
  \right.
\end{equation}
The table output is expressed in logarithmic format with 8-bit
fractional part, and thus $\zeta_f$ is given by
\begin{equation}
  \zeta_f = \frac{\ln{2}}{512} \simeq 1.4 \times 10^{-3}.
\end{equation}

\begin{figure}[hbtp]
\centerline{\psfig{file=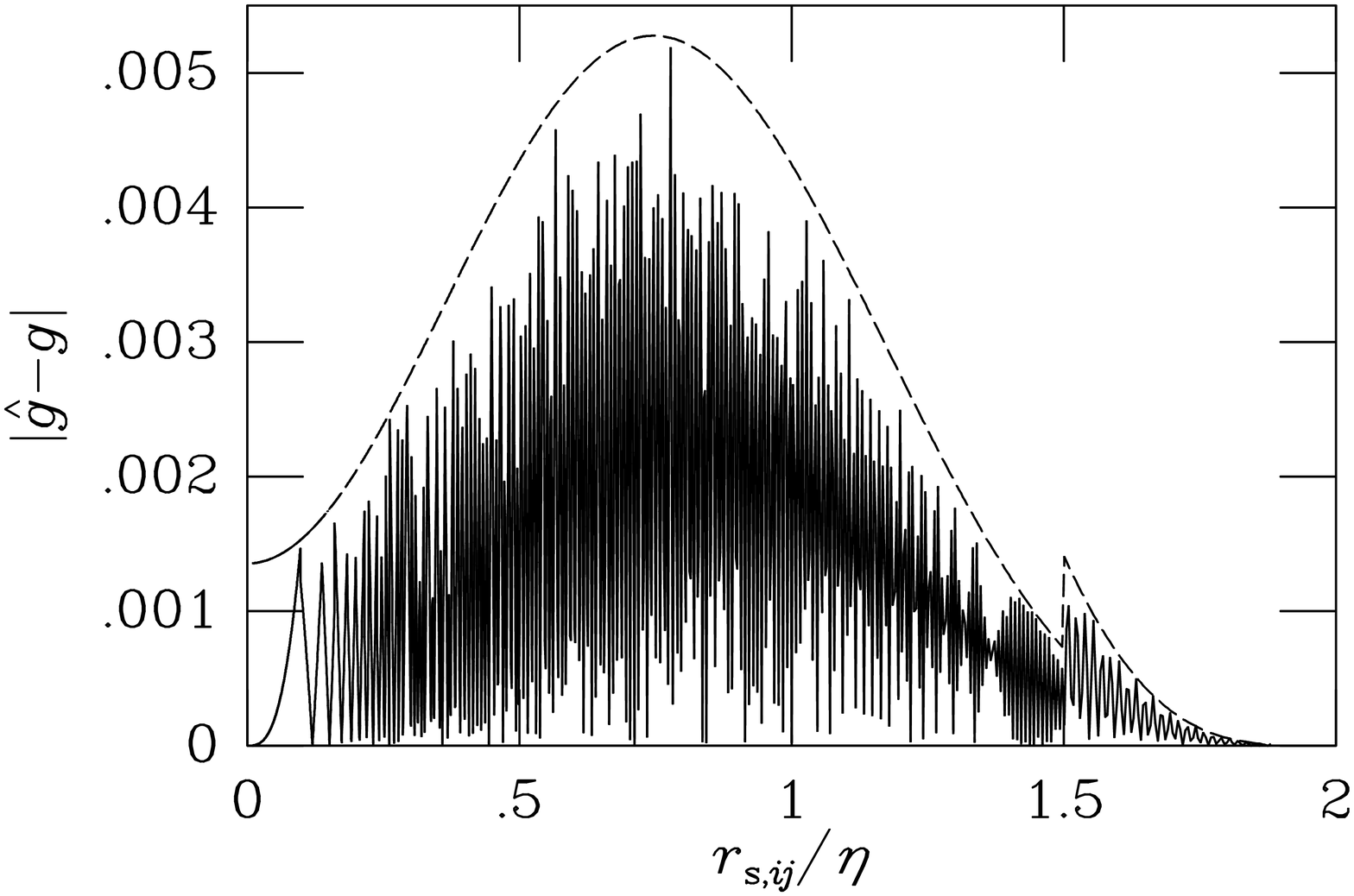,width=80mm}}
\caption{
Error of the cutoff function $g$ calculated by the cutoff function
table. Solid curve denotes measured error. Dashed curve is an estimate
of the error for the worst case. As the function $g$ we used $g_{\rm
P3M}$ for P${}^3$M method given by equation(\protect{\ref{eq:gp3m}}).
}
\label{fig:errg}
\end{figure}

Figure \ref{fig:errg} shows the theoretical and measured values of the
error as functions of $r_{{\rm s},ij}/\eta$. We used the cutoff
function for P${}^3$M method, $g_{\rm P3M}$, given by equation
(\ref{eq:gp3m}). The theoretical estimate and measured errors show
a good agreement. The error increases discontinuously at $r_{{\rm
s},ij}/\eta = 1.5$ because the resolution of the table entry degrades. 
Even though, the magnitude of the error is not more than 0.5\% in all
range.

In the following we estimate the error of the force with cutoff. We
define the error $S_r(\Vec{f}_{\rm c})$ as
\begin{equation}
  S_r(\Vec{f}_{\rm c}) \equiv
  \frac{\left<|\hat{\Vec{f}_{\rm c}}-\Vec{f}_{\rm c}|^2\right>}{f^2},
\end{equation}
where $\Vec{f}_{\rm c}$ and $\hat{\Vec{f}_{\rm c}}$ are exact and
calculated forces with cutoff, and $\Vec{f}$ is a force without
cutoff.  We defined the error relative to the force without cutoff,
since that is what affects the overall accuracy.

Following a similar procedure as that for the $1/r^2$ force, the force
calculation error is given by
\begin{eqnarray}\label{eq:pperrg}
  S_r(\Vec{f})
    & \lesssim &
        \left(
            \frac{18r^2}{r_{{\rm s},ij}^4} -
            \frac{12}{r_{{\rm s},ij}^2} +
            \frac{6}{r^2}
        \right)\epsilon_i^2 \nonumber \\
&&      + \left[
            \frac{9r^4}{r_{{\rm s},ij}^4} -
            \frac{6r^2}{r_{{\rm s},ij}^2} +
            \frac{17}{2} +
            \left(
                \frac{g'}{g}
                \frac{r_{{\rm s},ij}}{\eta}
            \right)^2
        \right]\epsilon_f^2 \nonumber \\
&&      + \left(
            \frac{g'}{g}
        \right)^2 \xi_i^2.
\end{eqnarray}
If $\epsilon \ll r$, the inequality becomes
\begin{eqnarray}
  S_r(\Vec{f})
      & \lesssim &
          \frac{12}{r^2} \epsilon^2_i
        + \left[
          \frac{23}{2} +
          \left(
              \frac{g'}{g}
              \frac{r}{\eta}
          \right)^2
        \right] \epsilon^2_f \\
&&       + \left(
            \frac{g'}{g}
        \right)^2 \xi_i^2,
\end{eqnarray}
where $\xi_i$ is the r.m.s. absolute error of the format used for the
entry of the cutoff function table, and is expressed as
\begin{equation}
  \xi_i = \frac{p}{256 \sqrt{3}}.
\end{equation}

\begin{figure}[hbtp]
\centerline{\psfig{file=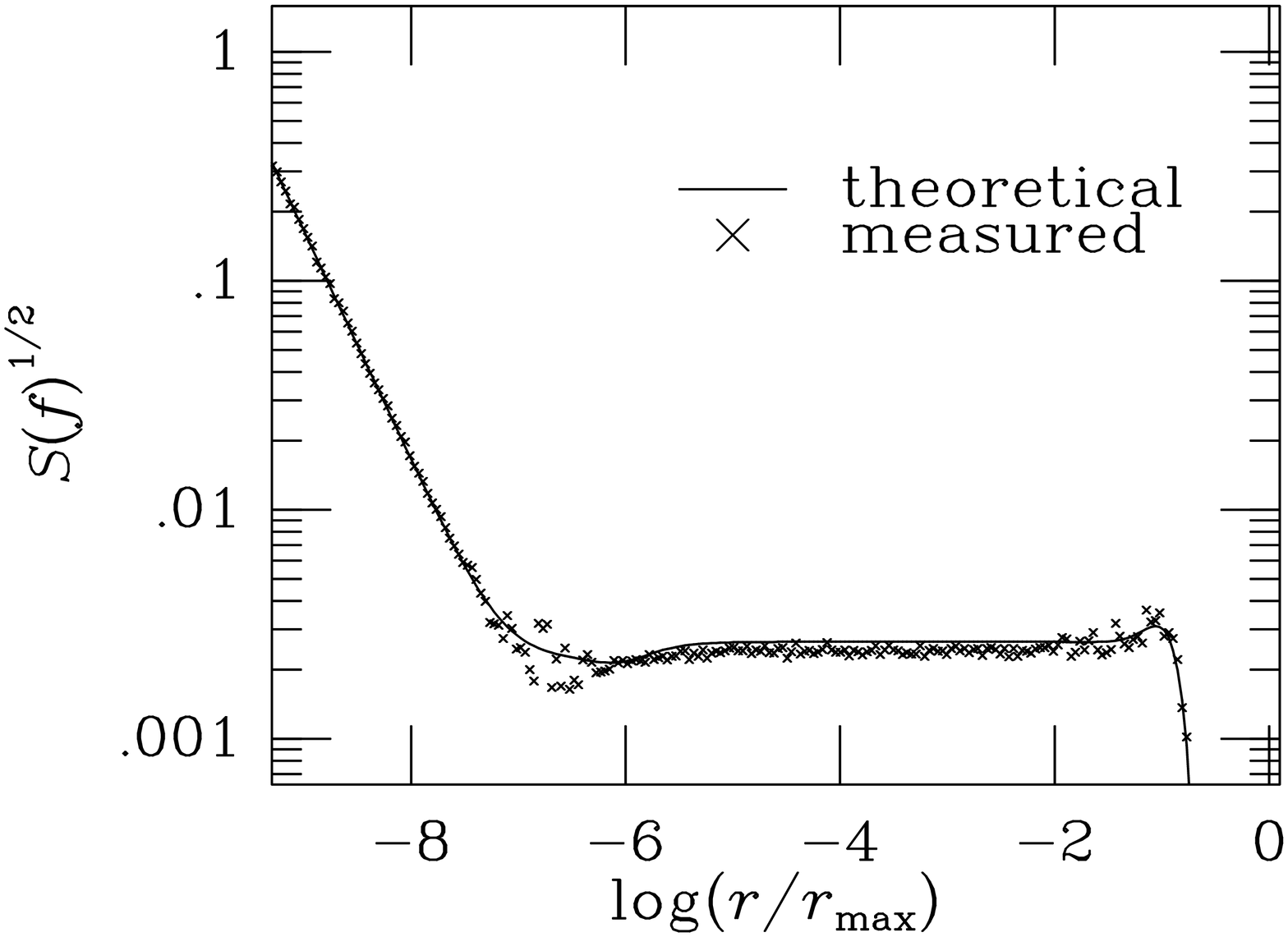,width=80mm}}
\caption{
Same as figure \protect{\ref{fig:err53}}, but for a force with cutoff
$g_{\rm P3M}$ for P${}^3$M method given by
equation(\protect{\ref{eq:gp3m}}).  The softening parameter $\epsilon$
and the scale length of the cutoff function $\eta$ are $ 10^{-6}
r_{\rm max}$ and $ r_{\rm max}/8$, respectively.
}
\label{fig:err5g}
\end{figure}

Figure \ref{fig:err5g} shows the theoretical and the measured error
$S(\Vec{f})^{1/2}$ for the force with cutoff $g_{\rm P3M}$. The
softening parameter is set to $\epsilon = 10^{-6} r_{\rm max}$.  The
scale length is set to $\eta = 1/8$. We can see that the error of the
force with cutoff is no more than 0.4\% for $r$ in the range of
$[10^{-6} r_{\rm max}, r_{\rm max}]$. Note that the measured error is
systematically better than the theoretical estimate of equation
(\ref{eq:pperrg}) for $10^{-6} r_{\rm max} < r < 10^{-2} r_{\rm max}$.
This is simply because in this range $g$ is very close to unity. The
output of the table is exactly one. Thus, the round-off error of the
cutoff table is smaller than the estimate of it.

\section{Timing Results}\label{sec:timing}

Here we give timing results of the GRAPE-5 system for direct and tree
algorithms. For P${}^3$M method, we give theoretical estimate only.
We used one processor board and a host computer COMPAQ AlphaStation
XP1000 (Alpha 21264 processor 500 MHz). For comparison, we also give
theoretical estimate for GRAPE-3 system. As the host computer of
GRAPE-3 system, we use SUN Ultra 2/200 (UltraSPARC processor 200 MHz).

\subsection{Direct Summation Algorithm}\label{sec:perfe}

The total calculation time per timestep is expressed as
\begin{equation}\label{eq:equt}
	T_{\rm eq} = T_{\rm host1} + T_{\rm grape1} + T_{\rm comm1},
\end{equation}
where $T_{\rm host1}$, $T_{\rm grape1}$, and $T_{\rm comm1}$ are the
time spent on the host computer, the time spent on GRAPE-5, and the
time spent for data transfer between the host computer and GRAPE-5,
respectively. The time spent on the host computer is expressed as
\begin{equation}
  \label{eq:equth}
  T_{\rm host1} = N t_{\rm misc1},\\
\end{equation}
where $t_{\rm misc1}$ represents calculation time spent on the host
computer per particle, which includes time integration of the
particles, diagnostics, and other miscellaneous $O(N)$ contributions.
The value of $t_{\rm misc1}$ and other timing constants used for
theoretical estimates in this section are summarized in table
\ref{tab:perf}.

\begin{table}[hbtp]
\caption{Timing constants used for performance estimation}\label{tab:perf}
\begin{center}
\begin{tabular}{lcc}
\hline
\hline
parameter & GRAPE-5 & GRAPE-3\\
          & (s)     & (s) \\
\hline
$t_{\rm misc1}$ & 1.1 $\times$ 10${}^{-6}$ & 5.4 $\times$ 10${}^{-6}$ \\
$t_{\rm pipe}$ & 1.25 $\times$ 10${}^{-8}$ & 5.0 $\times$ 10${}^{-8}$ \\
$t_{{\rm comm},i}$ & 2.1 $\times$ 10${}^{-8}$ & 3.3 $\times$ 10${}^{-7}$ \\
$t_{{\rm comm},j}$ & 3.3 $\times$ 10${}^{-8}$ & 3.3 $\times$ 10${}^{-7}$ \\
$t_{{\rm comm},f}$ & 3.6 $\times$ 10${}^{-8}$ & 3.3 $\times$ 10${}^{-7}$ \\
$t_{\rm con}$ & 4.2 $\times$ 10${}^{-7}$ & 2.0 $\times$ 10${}^{-6}$ \\
$t_{\rm list}$ & 5.9 $\times$ 10${}^{-7}$ & 2.8 $\times$ 10${}^{-6}$ \\
$t_{\rm misc2}$ & 8.5 $\times$ 10${}^{-7}$ & 4.0 $\times$ 10${}^{-6}$ \\
$t_{\rm fft}$ & 8.5 $\times$ 10${}^{-8}$ & - \\
$t_{\rm misc3}$ & 5.2 $\times$ 10${}^{-6}$ & - \\
\hline
\end{tabular}
\end{center}
\end{table}

The time spent on GRAPE-5 is expressed as
\begin{equation}
  \label{eq:equtg}
  T_{\rm grape1} = \frac{N^2 t_{\rm pipe}}{n_{\rm pipe}}.
\end{equation}
Here $n_{\rm pipe}$ is the number of real pipelines per processor
board, which equals 16. The number $t_{\rm pipe}$ is the cycle time of
the G5 chip.

The time spent for data transfer is expressed as
\begin{equation}
  \label{eq:equtc}
  T_{\rm comm1} = N (16 t_{{\rm comm},j} +
                     12 q t_{{\rm comm},i} +
                     32 q t_{{\rm comm},f}),
\end{equation}
where $t_{{\rm comm},j}$ and $t_{{\rm comm},i}$ are the time to
transfer one byte data from the host computer to GRAPE-5 for particle
$j$ and particle $i$, respectively. The transfer speed for particle
$j$ is faster than that for particle $i$, since the size of the data
transferred in one transaction is larger for particle $j$. Large data
implies small overhead and fast overall speed. The constant $t_{{\rm
comm},f}$ is the time to transfer one byte data from GRAPE-5 to the
host computer for calculated force, and $q$ is given by
\begin{equation}
	\label{eq:q1}
	q = \left\lfloor \frac{N-1}{n_{\rm mem}} + 1 \right\rfloor,
\end{equation}
where $\lfloor x \rfloor$ denotes the maximum integer which does not
exceed $x$, and $n_{\rm mem} (= 131072)$ is the maximum number of
particles which can be stored in the memory unit of a GRAPE-5
processor board. This $q$ indicates how the total force on a particle
is divided. If $N > n_{\rm mem}$, the force on a particle must be
divided into $q$ pieces, and GRAPE-5 must be used $q$ times to obtain
the total force.

\begin{figure}[hbtp]
\centerline{\psfig{file=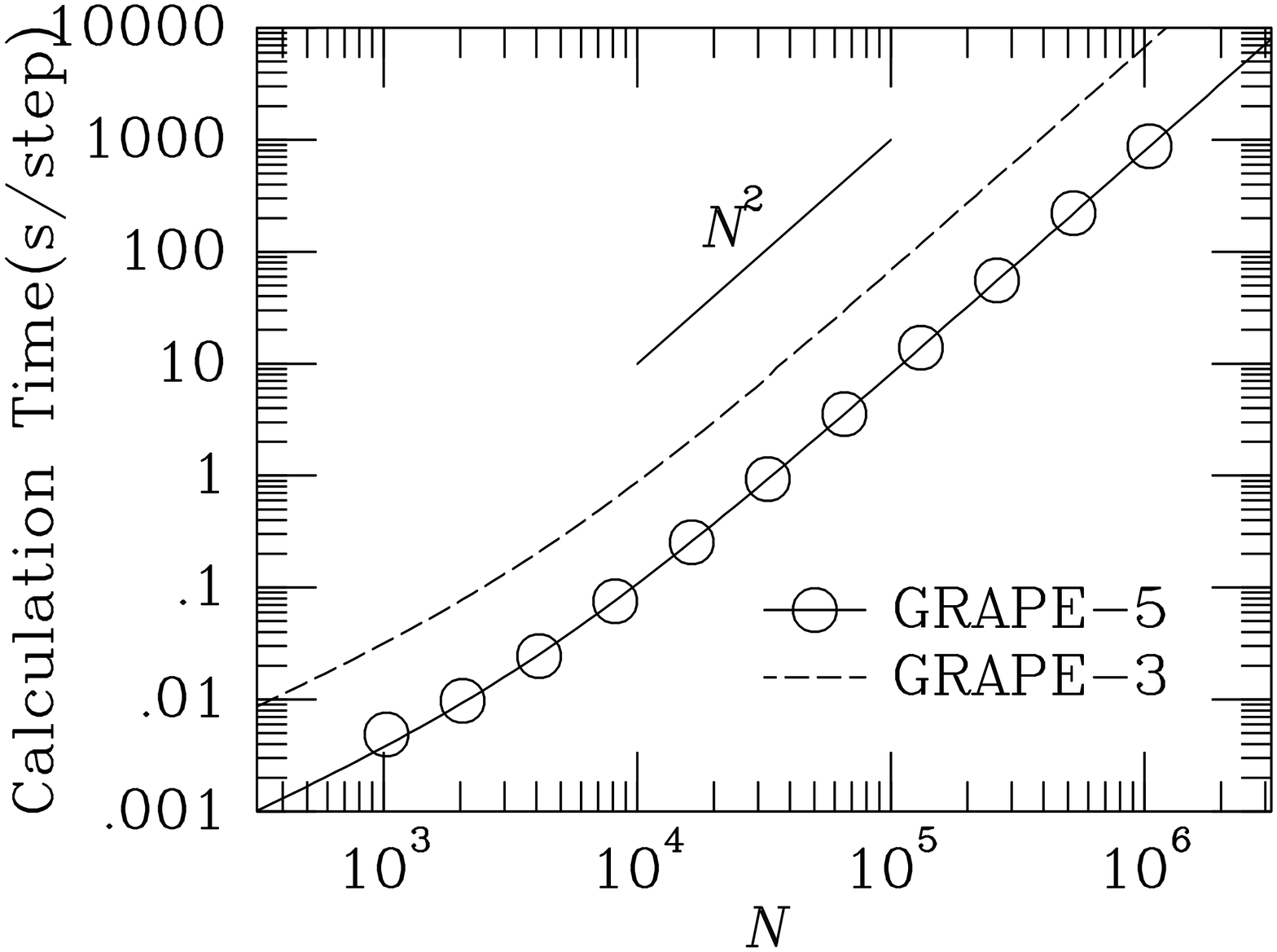,width=80mm}}
\caption{
Calculation time per one timestep for direct summation algorithm,
plotted as functions of the number of particles $N$.  The solid and
dashed curves are theoretical estimate for GRAPE-5 and GRAPE-3,
respectively. The open circles represent the measured performance of
GRAPE-5.
}
\label{fig:perfeq}
\end{figure}

\begin{figure}[hbtp]
\centerline{\psfig{file=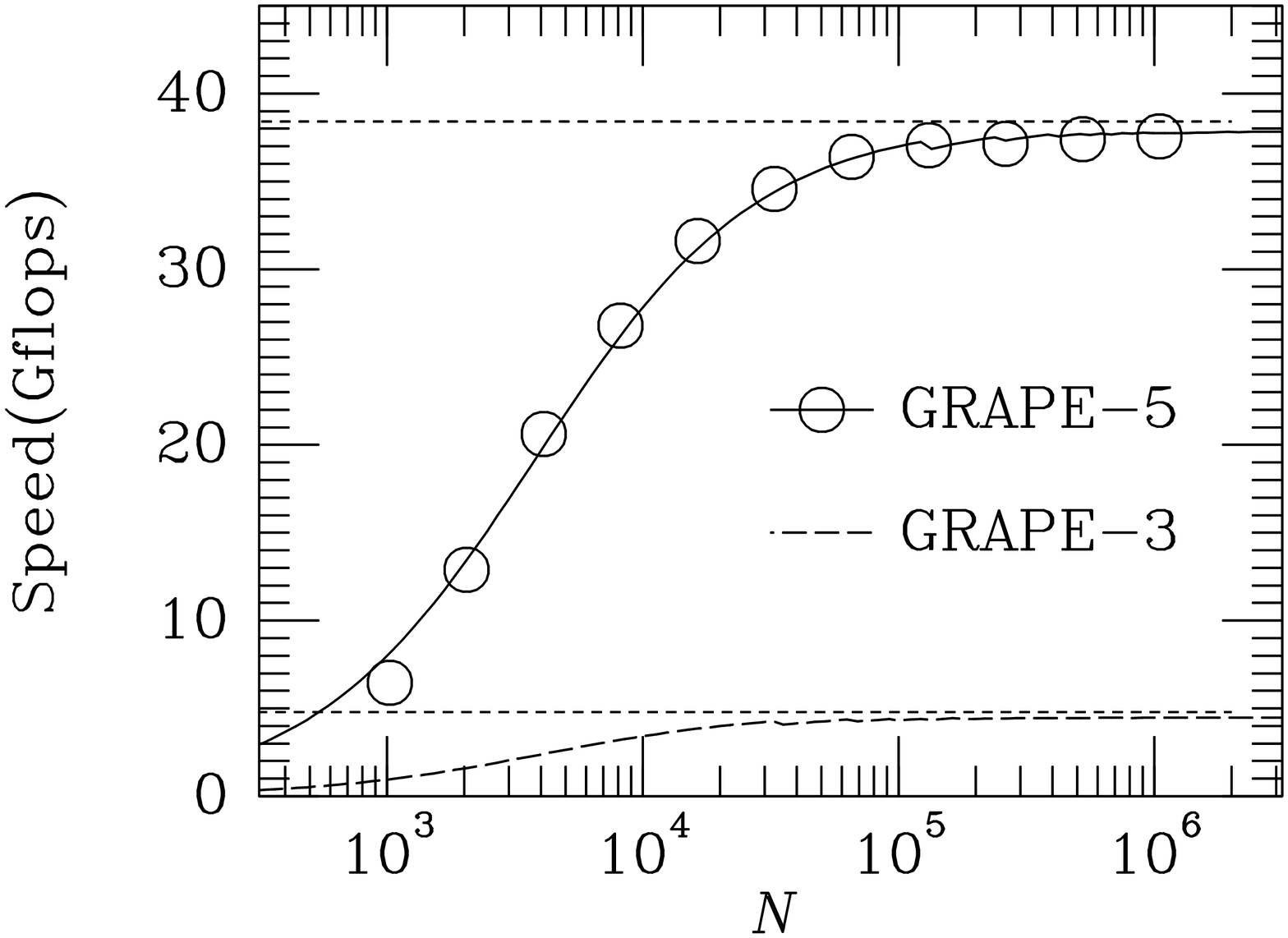,width=80mm}}
\caption{
Calculation speed for direct summation algorithm, plotted as a
function of the number of particles $N$.  The solid and dashed curves
are theoretical estimate for GRAPE-5 and GRAPE-3, respectively.  The
open circles represent the measured performance of GRAPE-5.  The
dotted lines denote peak speed of GRAPE-5 and GRAPE-3.
}
\label{fig:effieq}
\end{figure}

Figure \ref{fig:perfeq} shows measured and estimated calculation time
per one timestep as functions of $N$. Figure \ref{fig:effieq} shows
measured and estimated speed as functions of the number of particle
$N$.
For GRAPE-5 system, we chose timing constant of the host computer,
$t_{\rm host1}$ to fit the measured results. For GRAPE-3 system, we
used $t_{\rm host1}$ scaled from the one for GRAPE-5 system, according
to the ratio of SPECfp95 values of the hosts. For the data transfer
times, we used measured values for GRAPE-5 system, and the value given
in Athanassoula {\it et al.} (1998) for GRAPE-3 system.


In figure \ref{fig:perfeq}, we can see that the calculation with
GRAPE-5 is about 8 times faster than that with GRAPE-3, for the entire
range of $N$ we used.  In figure \ref{fig:effieq}, we can see that the
effective performance of GRAPE-5 exceeds 70 \% of the peak
performance, for rather small $N$ ($= 10^4$).

\subsection{Barnes-Hut Tree Algorithm}\label{sec:perft}

The total calculation time per timestep is expressed as
\begin{equation}
	\label{eq:treet}
	T_{\rm tree} = T_{\rm host2} + T_{\rm grape2} + T_{\rm comm2},
\end{equation}
where $T_{\rm host2}$, $T_{\rm grape2}$, and $T_{\rm comm2}$ are the
time spent on the host computer, the time spent on GRAPE-5, and the
time spent for data transfer between the host computer and GRAPE-5,
respectively. The time spent on the host computer is expressed as
\begin{equation}
	\label{eq:treeth}
	T_{\rm host2} =
      (N \log_{10} N) t_{\rm con} +
      \frac{N n_{\rm terms}}{n_{\rm g}} t_{\rm list} +
      N t_{\rm misc2},
\end{equation}
where $t_{\rm con}$ is the time to construct the tree structure,
$t_{\rm list}$ is the time to construct the interaction lists, and
$t_{\rm misc2}$ represents $O(N)$ miscellaneous contributions
(Fukushige {\it et al.} 1991).  In this equation, $n_{\rm terms}$ is
the average length of the interaction list.  According to Makino
(1991), $n_{\rm terms}$ is estimated as follows:
\begin{eqnarray}
	\label{eq:nt}
n_{\rm terms} & \simeq & n_{\rm g} +
		14n_{\rm g}^{2/3} +
		84n_{\rm g}^{1/3} +
		56\log_8 n_{\rm g}  \nonumber \\
& &		 - 31 \theta^{-3} \log_{10}n_{\rm g} - 72  \nonumber \\
& &		+ 10^2 \theta^{-3} \log_{10} {\frac{N\theta^3}{23}}.
\end{eqnarray}
Here, $\theta$ is the opening angle.

The time spent on GRAPE-5 is expressed as
\begin{equation}
	\label{eq:treetg}
	T_{\rm grape2} = \frac{N n_{\rm terms} t_{\rm pipe}}{n_{\rm pipe}},\\
\end{equation}
and the time spent for data transfer is expressed as
\begin{equation}
	\label{eq:treetc}
	T_{\rm comm2} = 
      N \left( 16 \frac{n_{\rm terms}}{n_{\rm g}}t_{{\rm comm},j} +
               12 t_{{\rm comm},i} +
               32 t_{{\rm comm},f} \right).
\end{equation}

\begin{figure}[hbtp]
\centerline{\psfig{file=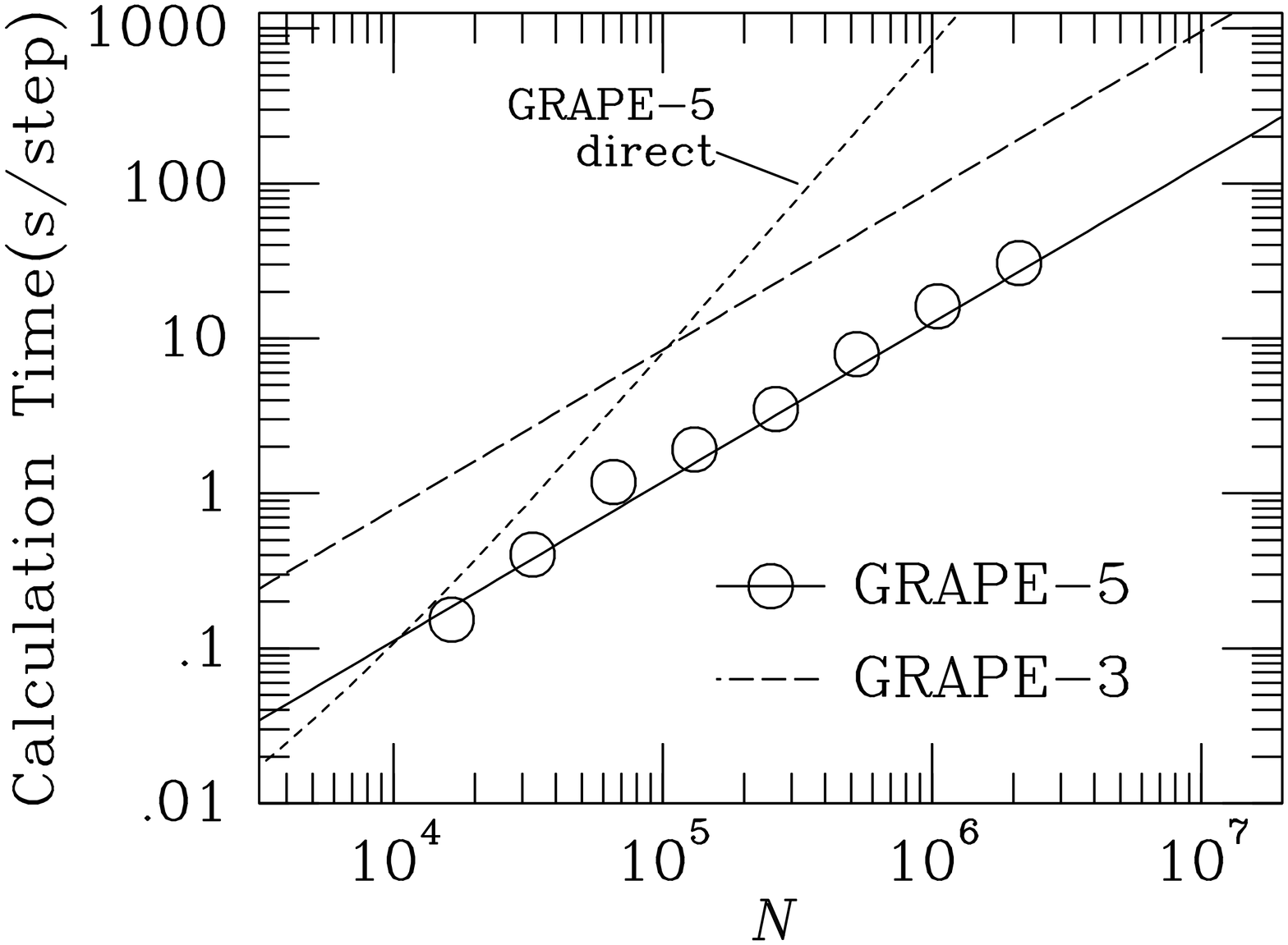,width=80mm}}
\caption{
Same as figure \protect{\ref{fig:perfeq}}, but for the Barnes-Hut tree
algorithm.
}
\label{fig:perftree}
\end{figure}

Figure \ref{fig:perftree} shows the calculation time per one timestep. 
Estimates for GRAPE-3 are also shown. Table \ref{tab:fractree} shows
the calculation time spent on GRAPE-5, on the host computer, and for
data transfer for the number of particle $N$=1048576. As the initial
particle distribution, we use a Plummer model.  All particles have
equal mass. The opening angle $\theta$ is 0.75.  For GRAPE-5, $n_{\rm
g} \simeq 2000$. For GRAPE-3, we used $n_{\rm g} \simeq 1000$. The
timing constants for the host computer and data transfer are chosen in
the same way as in section \ref{sec:perfe}.

\begin{table}[hbtp]
\caption{Calculation time per one timestep for Barnes-Hut tree algorithm
(Plummer model, $N$=1048576, $\theta$=0.75)}\label{tab:fractree}
\begin{center}
\begin{tabular}{lcc}
\hline
\hline
                       &  & Estimate for \\
                       & GRAPE-5 & GRAPE-3 \\
                       & (s)     & (s) \\
\hline
Host computer         & & \\
~~Tree construction & 2.8      & 13.0 \\
~~Interaction list construction \hspace*{-5mm} & 2.0      & 12.7 \\
~~Miscellaneous     & 1.3      & 4.4 \\
GRAPE                 & 6.9      & 28.4\\
Data transfer         & 3.1      & 40.3 \\
Total                 & 16.1     & 98.8 \\
\hline
\end{tabular}
\end{center}
\end{table}

In figure \ref{fig:perftree}, we can see that the calculation with
GRAPE-5 is about 6 times faster than that with GRAPE-3, and that the
tree algorithm is faster than the direct summation algorithm for $N
\gtrsim 10^4$. Table \ref{tab:fractree} indicates that we can use up to
two or three processor boards to increase the calculation speed
further. In order to use more than four boards effectively, we need a
host computer which has faster computing speed and multiple
communication buses, so that the calculation on the host computer and
the data transfer do not limit the total performance.

The calculation time with single processor board for a million
particle simulation is 16 seconds per timestep. To put this number
into perspective, parallel treecode by Dubinski (1996) would took
around 60 seconds for one million particles on 64 processor T3D, with
$\theta = 1.2$. Taking into account the difference in $\theta$, it is
perhaps fair to say our treecode on GRAPE-5 is about 6 times faster
than Dubinski's parallel code on 64 processor T3D (his code does not
scale well for more than 128 processors).  Yahagi {\it et al.} (1999)
described an implementation of treecode on Fujitsu VPP300/16R, which
took $\sim 7$ seconds for one million particles, with the opening
criterion roughly corresponding to $\theta = 1$. Again taking into
account the difference in $\theta$, the effective speed of our code is
$\sim 70 \%$ of VPP300/16R.

\subsection{P${}^3$M Method}\label{sec:perfp}

The total calculation time per timestep is expressed as
\begin{equation}
	\label{treet}
	T_{\rm p3m} = T_{\rm host3} + T_{\rm grape3} + T_{\rm comm3},
\end{equation}
where $T_{\rm host3}$, $T_{\rm grape3}$, and $T_{\rm comm3}$ are the
time spent on the host computer, the time spent on GRAPE-5, and the
time spent for data transfer between the host computer and GRAPE-5,
respectively. In the following we present the estimate for the case of 
homogeneous distribution of particles.

The time spent on the host computer is expressed as
\begin{equation}
	\label{eq:p3mth}
	T_{\rm host3} = 3 M_{\rm pm} \log_2 M_{\rm pm} t_{\rm fft} +N t_{\rm misc3}.
\end{equation}
The first term represents the time for FFT and the second term
represents the time for $O(N)$ miscellaneous calculations. Here
$M_{\rm pm}$ is the number of meshes in one-dimension used in PM force
calculation, and $t_{\rm misc3}$ represents time for miscellaneous
calculation per particle.

The time spent on GRAPE-5 is expressed as
\begin{equation}
	\label{eq:p3mtg}
	T_{\rm grape3} = \frac{27 N^2 t_{\rm pipe}}{M_{\rm pp}^{3} n_{\rm pipe}},
\end{equation}
where $M_{\rm pp}$ is the number of meshes for PP force. We set the
ratio $M_{\rm pm}/M_{\rm pp}$ to 2.9, following Brieu {\it et
al.} (1995) and Fukushige {\it et al.} (1996).
We set $M_{\rm pm}$ so as to minimize the total calculation time. The
optimal value of $M_{\rm pm}$ is given approximately by
\begin{equation}
  M_{\rm pm}^3 \simeq 11 \sqrt{
                      \frac{t_{\rm pipe}}{n_{\rm pipe} t_{\rm fft}}} N.
\end{equation}

The time spent for data transfer is expressed as
\begin{equation}
	\label{eq:p3mtc}
    T_{\rm comm3} = N (16 t_{{\rm comm},j} + 
                       12 t_{{\rm comm},i} +
                       32 t_{{\rm comm},f}).
\end{equation}

\begin{figure}[hbtp]
\centerline{\psfig{file=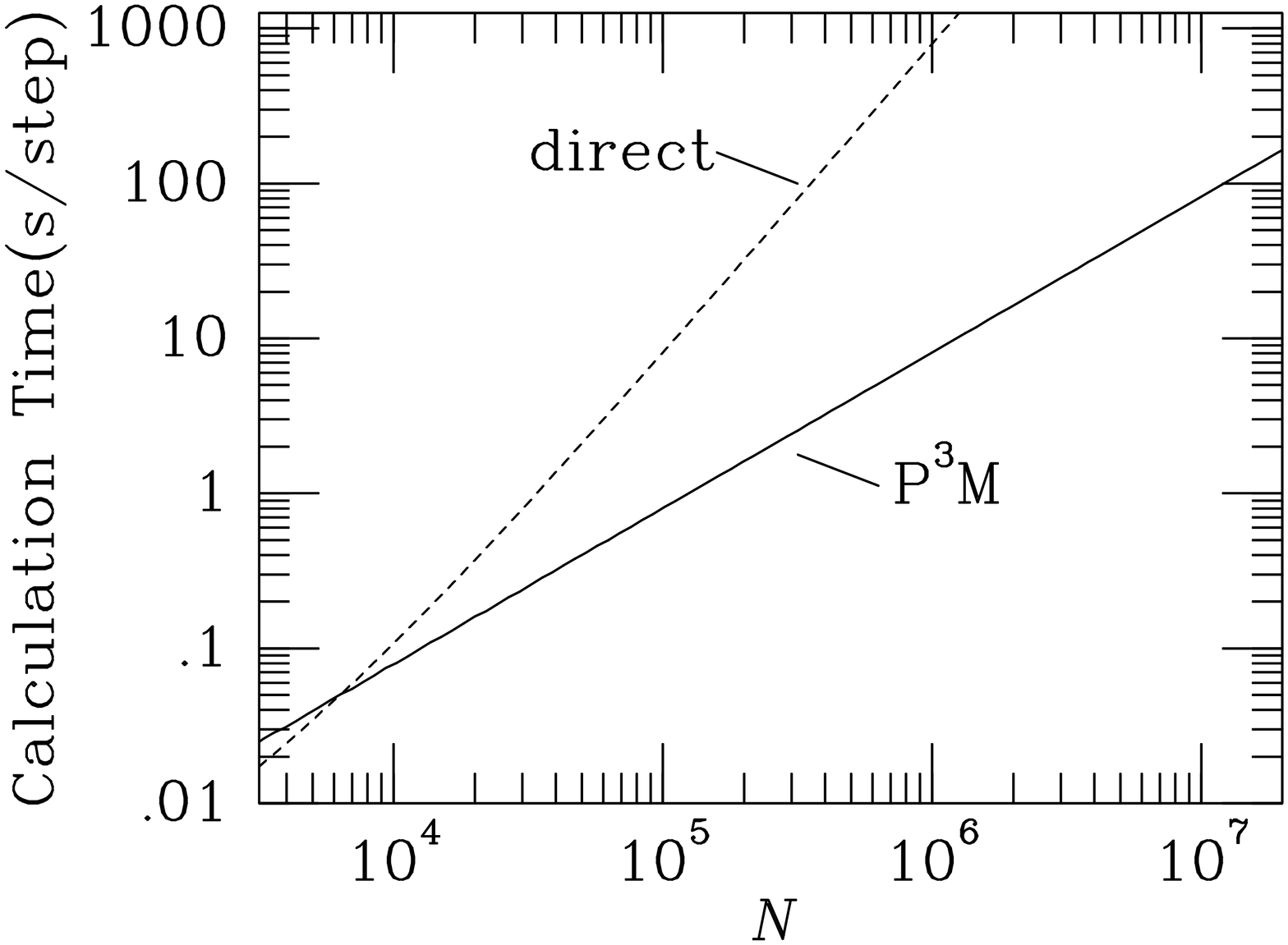,width=80mm}}
\caption{
Same as figure \protect{\ref{fig:perfeq}}, but for the P${}^3$M
method. Actual measurement is not available.
}
\label{fig:perfp3m}
\end{figure}

Figure \ref{fig:perfp3m} shows the calculation time per one timestep.
Theoretical estimate for GRAPE-5 is plotted. As the timing constants
for the host computer, $t_{\rm fft}$ and $t_{\rm misc3}$, we used
values scaled from the ones given in Fukushige {\it et al.} 1996,
according to the SPECfp95 values of the hosts.

The estimated calculation speed is 40 and 25 times faster than that
with GRAPE-3 (Brieu {\it et al.} 1995) and MD-GRAPE (Fukushige {\it et
al.} 1996), respectively. The calculation time with single processor
board for 16 million particles simulation is 150 seconds per timestep.
As discussed by Brieu {\it et al.} (1995), this calculation time
depends only weakly on the degree of the clustering. This is simply
because the calculation cost of $O(N)$ part of the code (mass
assignment, particle update, etc.) dominates the total cost. Even at a
highly clustered stage, the increase in the calculation cost would be a
factor of two or so. On the other hand, CPU time of parallel P${}^3$M
implementation is very sensitive to clustering. For example,
MacFarland {\it et al.} (1998) reported the CPU time per timestep
increased from $\sim 10$ seconds to $\ge 100$ seconds for their 16
million particle simulation on 128 processor T3E-600. Thus, depending
on the degree of clustering, P${}^3$M on GRAPE-5 runs at the speed of
5-50 \% of a 128 processor T3E-600.

\section{Future Prospects}\label{sec:future}

\subsection{Massively-Parallel GRAPE-5}

We plan to build a massively-parallel GRAPE-5 system with peak
performance of about 1 Tflops. This system will consist of about 20
processor boards and a host computer. Each processor board is
connected to a PHIB which is inserted into one PCI slot. The host
computer has multiple processors and multiple PCI slots.

In order for 1 Tflops-peak GRAPE-5 system to operate efficiently with
the tree algorithm, a host computer should have the effective
computing speed and communication speed of about 5 Gflops and a few
hundreds MB/s, respectively. Currently, these performance figures are
offered only by machines with multiple processors.

We consider two types of host computers. One is a shared-memory
multiprocessors (SMPs), and the other is a workstation cluster.

The advantage of the SMP over the workstation cluster is that the
implementation of the code is relatively easy. The disadvantages are
that the price is relatively high and the number of processors is
limited to 10 or smaller. This limit is due to bottleneck in the
memory access from multiple processors.

The advantage of the workstation cluster is that they are inexpensive
and that it is possible to connect $\sim$ 100 processors. For example,
Warren {\it et al.}(1998) performed cosmological $N$-body simulation
with tree algorithm on Avalon, a cluster of 70 DEC Alpha processors
(DEC Alpha 21164A, 533 MHz). The disadvantage is that the
implementation of the code is more difficult.

We plan to build a system based on a cluster of 4-8 workstations. A
preliminary analysis indicates that 100BT Ethernet connection offers
sufficient communication bandwidth. On this system, one timestep of
10${}^7$-body simulation with tree algorithm would take about 10
seconds. Using this system, for example, we can perform cosmological
10${}^8$-body simulation for 10${}^3$ steps in one day.

\subsection{GRAPE-5/PROGRAPE System}

We plan to construct a heterogeneous computing system,
GRAPE-5/PROGRAPE system. PROGRAPE (PROgrammable GRAPE; Hamada {\it et
al.} 1999) is a multi-purpose computer for many-body simulation. It
consists of reconfigurable processor implemented on FPGA (Field
Programmable Gate Array) chips and memory which stores particle
data. PROGRAPE has a very similar architecture to GRAPE except that it
uses FPGA chips as pipeline processor instead of custom LSI chips such
as G5 chip. PROGRAPE can calculate any interaction which is expressed
as
\begin{equation}
  \Vec{f}_{i} = \sum_j \Vec{g}(\Vec{p}_i, \Vec{p}_j),
\end{equation}
where $i$ and $j$ are the indices to the two sets of particles,
$\Vec{p}_i$ is the data of particle $i$ (for example position and
mass, but can include quantities such as radius, pressure,
temperature, etc.), and $\Vec{g}(\Vec{p}_i, \Vec{p}_j)$ is an
arbitrary function of these particle data.

Hamada {\it et al.} (1999) have developed the first prototype of
PROGRAPE, PROGRAPE-1. It houses two FPGA devices (EPF10K100, Altera
Corp.), each of which has 100k logical gates. They have implemented
gravitational pipelines same as that in GRAPE chip, which are used for
GRAPE-3. The pipelines operates at a clock cycle of 16 MHz and the
peak performance is 0.96 Gflops.

One application example of the GRAPE-5/PROGRAPE system is Smoothed
Particle Hydrodynamics (SPH; Lucy 1977, Monaghan 1985). SPH is widely
used in gas dynamical simulations in astrophysics. GRAPE-5 calculates
the gravitational interactions and PROGRAPE calculates hydrodynamical
interactions, which includes calculations of density, pressure
gradient, and artificial viscosity. For the more detail about PROGRAPE
system, see Hamada {\it et al.} (1999).

Another example is a simulation with Ewald method (Ewald 1921). The
Ewald method is a method to calculate gravitational force under
periodic boundary condition. In the Ewald method, the force is divided
into two interactions in real space and in wave number space. GRAPE-5
calculates the interaction in real space and PROGRAPE calculates the
interaction in wave-number space.

It is difficult to estimate the exact performance of GRAPE-5/PROGRAPE
system, since the performance of PROGRAPE varies depending on
applications. But we can expect that the above mentioned simulations
on GRAPE-5/PROGRAPE system would be faster than that on GRAPE-5 system
without PROGRAPE by at least a factor of 10.

\acknowledgements

We would like to thank Daiichiro Sugimoto, who have started the GRAPE
project. This work was partially supported by the Research for the
Future Program of Japan Society for the Promotion of Science,
JSPS-RFTP 97P01102.

\end{document}